\documentclass[useAMS,usenatbib]{mnras}

\usepackage{epsfig}
\usepackage{aas_macros}
\usepackage{natbib}
\usepackage{times}
\usepackage{graphicx, bm, amssymb, xcolor}
\usepackage{verbatim}
\usepackage[all]{hypcap}
\usepackage{xspace}
\usepackage{multirow}
\usepackage{pdflscape}
\usepackage{pgf,tikz}
\usepackage{mathrsfs}
\usetikzlibrary{arrows}
\usepackage{amsmath}
\usepackage{subfigure}
\usepackage{array}
\usepackage{algorithm}
\usepackage[noend]{algpseudocode}

\usepackage{environ}
\makeatletter
\newsavebox{\measure@tikzpicture}
\NewEnviron{scaletikzpicturetowidth}[1]{%
 \def\tikz@width{#1}%
 \begin{lrbox}{\measure@tikzpicture}%
 \BODY
 \end{lrbox}%
 \pgfmathparse{#1/\wd\measure@tikzpicture}%
 \BODY
}
\makeatother

\newcommand{\be}{\begin{equation}}
\newcommand{\ee}{\end{equation}}
\newcommand{\bea}{\begin{eqnarray}}
\newcommand{\eea}{\end{eqnarray}}
\newcommand{\dd}{\mathrm{d}}

\title{A scaling relation for the molecular cloud lifetime in Milky Way-like galaxies}
\author[]{Sarah~M.~R.~Jeffreson\thanks{sarah.jeffreson@cfa.harvard.edu}$^{1,2}$, Benjamin~W.~Keller$^{1}$, Andrew~J.~Winter$^{3}$,
M{\'e}lanie~Chevance$^{1}$,
\newauthor
J.~M.~Diederik~Kruijssen$^{1}$, Mark~R.~Krumholz$^{4}$, Yusuke Fujimoto$^{5}$ \\
\\
$^{1}$ Astronomisches Rechen-Institut, Zentrum f\"{u}r Astronomie der Universit\"{a}t Heidelberg, M\"{o}nchhofstra\ss e 12-14, 69120 Heidelberg, Germany \\
$^{2}$ Center for Astrophysics, Harvard \& Smithsonian, 60 Garden St, Cambridge, MA 02138, USA \\
$^{3}$ Institut f\"{u}r Theoretische Astrophysik, Zentrum f\"{u}r Astronomie der Universit\"{a}t Heidelberg, Albert-Ueberle-Str.~2, 69120 Heidelberg, Germany \\
$^{4}$ Research School of Astronomy and Astrophysics, Australian National University, Canberra, ACT 2611 Australia \\
$^{5}$ Earth and Planets Laboratory, Carnegie Institution for Science, 5241 Broad Branch Road, NW, Washington, DC 20015, USA \\
}

\hypersetup{draft}
\begin{document}


\pagerange{\pageref{firstpage}--\pageref{lastpage}} \pubyear{2020}

\maketitle

\label{firstpage}

\begin{abstract}
We study the time evolution of molecular clouds across three Milky Way-like isolated disc galaxy simulations at a temporal resolution of $1$~Myr, and at a range of spatial resolutions spanning two orders of magnitude in spatial scale from $\sim 10$~pc up to $\sim 1$~kpc. The cloud evolution networks generated at the highest spatial resolution contain a cumulative total of $\sim 80,000$ separate molecular clouds in different galactic-dynamical environments. We find that clouds undergo mergers at a rate proportional to the crossing time between their centroids, but that their physical properties are largely insensitive to these interactions. Below the gas disc scale-height, the cloud lifetime $\tau_{\rm life}$ obeys a scaling relation of the form $\tau_{\rm life} \propto \ell^{-0.3}$ with the cloud size $\ell$, consistent with over-densities that collapse, form stars, and are dispersed by stellar feedback. Above the disc scale-height, these self-gravitating regions are no longer resolved, so the scaling relation flattens to a constant value of $\sim 13$~Myr, consistent with the turbulent crossing time of the gas disc, as observed in nearby disc galaxies.
\end{abstract}

\begin{keywords}
stars: formation --- ISM: clouds --- ISM: evolution --- ISM: kinematics and dynamics --- galaxies: evolution --- galaxies: ISM
\end{keywords}

\section{Introduction}
\label{Sec::Introduction}
Giant molecular clouds provide the reservoirs of cold molecular gas from which the majority of stars are formed~\citep{Kennicutt+Evans12}. Their lifetimes place an upper bound on the time-scale for star formation at a given spatial scale, which in combination with observations of the gas and star formation rate surface densities~\citep[e.g.][]{Kennicutt98,Bigiel08,Leroy08,Blanc09,Schruba10,Liu11}, constrains the value of the local star formation efficiency (SFE). As such, a prediction for the molecular cloud lifetime provides two key insights about the conversion of gas to stars in galaxies: (1) an understanding of the physics that limit the duration of star-formation episodes, and (2) a prediction for the fraction of gas that is converted to stars during these episodes.

Over the past two decades, observational evidence has mounted to support the view of molecular clouds as rapidly-evolving entities with lifetimes of order the dynamical time-scale, varying between $10$ and $55$~Myr~\citep{Engargiola03,Blitz2007,Kawamura09,Murray11,2012ApJ...761...37M,Meidt15,Corbelli17,Kruijssen2019,Chevance20,ChevanceReview20}. These measurements contrast with past studies that tie molecular cloud lifetimes to the $\ga 100$~Myr survival times of their constituent ${\rm H}_2$ molecules~\citep[e.g.][]{ScovilleHersh1979,1975ApJ...199L.105S,Koda09}. To complement studies of giant molecular cloud time-scales, a growing body of observational evidence now points towards a correlation of cloud properties with the galactic environment, across a range of spatial scales. In particular, significant environmental variation has been found in the gas depletion time~\citep{Leroy08}, the molecular cloud surface density, turbulent velocity dispersion and turbulent pressure~\citep[e.g.][]{Hughes13b,Sun18,Colombo+19,Sun2020}, the molecular cloud size~\citep{Heyer+09,Roman-Duval+10,Rice+16,Miville-Deschenes17,Colombo+19}, the molecular cloud mass~\citep{Colombo14,Hughes2016,Freeman17}, the galactic dense gas fraction~\citep{Usero15,Bigiel16}, and possibly (depending on the assumed CO to mass conversion factor) the SFE per free-fall time~\citep{Utomo18,2019ApJ...883....2S}. In accordance with both numerical simulations~\citep{TaskerTan09,DobbsPringle13,2014MNRAS.439..936F,Dobbs15,Benincasa2019,Jeffreson20} and analytic predictions~\citep{2015A&A...580A..49I,2017ApJ...836..175K,Meidt18,Jeffreson+Kruijssen18}, observed molecular clouds do not form and evolve in isolation, but are part of a network of galactic processes spanning from the kpc-scales of galactic dynamics down to the sub-cloud physics of star formation and stellar feedback.

Within this hierarchical baryon cycle, it has long been known that the time-scales associated with star formation vary as a function of spatial scale~\citep[e.g.][]{1996ApJ...466..802E,1998MNRAS.299..588E,2000ApJ...530..277E} according to the hierarchical~\citep[e.g.][]{1985prpl.conf..201S,1987ApJS...65...13B,1990ASSL..162..151S,1990ApJ...355..536L,1991ApJ...378..186F,1991IAUS..147...11B,1996ApJ...471..816E,2009A&A...507..355F} and supersonically-turbulent~\citep[e.g.][]{1994ApJ...423..681V,1995ApJ...455..536P,1997ApJ...474..730P,1998PhRvE..58.4501P,1998ApJ...508L..99S,2001ApJ...546..980O,KimWT&Ostriker03,2009ApJ...692..364F,2010A&A...512A..81F,2020NatAs...4.1064H} structure of the interstellar medium. In recent years, theories of star formation have begun to explore the spatial dependence of empirical star formation relations~\citep{2012ApJ...745...69K,2013ApJ...779....8K,Kruijssen2014,Semenov17,Semenov18,Semenov19,2019MNRAS.487.3845C,2020MNRAS.497..698T}, and observations have shifted towards the characterisation of molecular gas properties as an explicit function of spatial resolution~\citep{2013ApJ...769L..12L,Sun18,Schinnerer19,Sun2020}. Such analyses describe the duration and efficiency of galactic-scale star formation with respect to the physics driving the hierarchy of sub-galactic time-scales for molecular gas evolution.

In this work, we explore the time evolution of molecular cloud populations in Milky Way-mass galaxies as a function of spatial scale, using a set of three isolated galaxy simulations spanning a wide range of galactic-dynamical environments~\citep{Jeffreson20}. We construct detailed cloud evolution networks spanning over two orders of magnitude in spatial resolution, allowing us to probe the time-evolution and star-forming behaviour of molecular gas across a range of hierarchical levels in the interstellar medium. We compute the characteristic molecular cloud lifetime and cloud merger rate as a function of spatial scale, and examine how these quantities relate to the time-scales for star formation and gravitational collapse. Finally, we connect the derived scaling relations, where possible, to the galactic-dynamical environment and its influence (or lack thereof) on the clouds in our sample.

The remainder of this paper is structured as follows. In Section~\ref{Sec::simulations}, we re-iterate the key details of the three isolated galaxy simulations presented in~\cite{Jeffreson20}. Section~\ref{Sec::construct-cloud-history} describes how these simulations are used to construct the detailed cloud evolution networks analysed in this work. In Sections~\ref{Sec::fractal-structure} and~\ref{Sec::cloud-lifetime} we report the key results of our analysis: the spatial scalings of the cloud merger rate and the characteristic molecular cloud lifetime, respectively. Section~\ref{Sec::discussion} presents a discussion of our results in the context of existing simulations and observations of giant molecular cloud lifetimes and mergers. Finally, a summary of our conclusions is given in Section~\ref{Sec::conclusion}.

\section{Simulations} \label{Sec::simulations}
\begin{figure*}
 \label{Fig::morphology}
  \includegraphics[width=\linewidth]{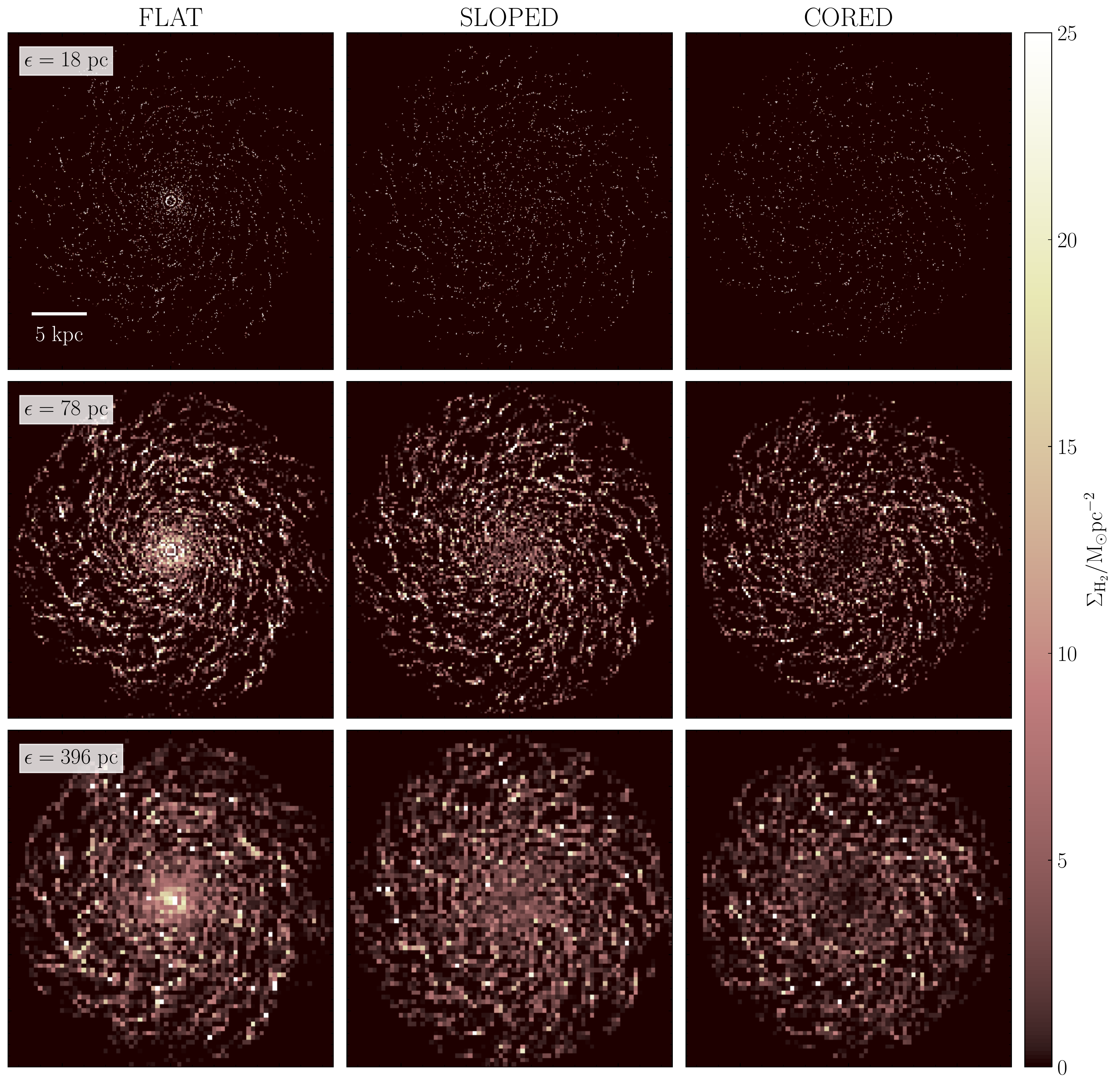}
  \caption{Column density maps of the molecular gas in each simulation, down-sampled from the native map resolution of $\epsilon = 6$~pc to spatial resolutions of $\epsilon = 18$~pc (top row), $78$~pc (central row) and $396$~pc (bottom row).}
\end{figure*}
\begin{figure}
 \label{Fig::rotcurve-components}
  \includegraphics[width=\linewidth]{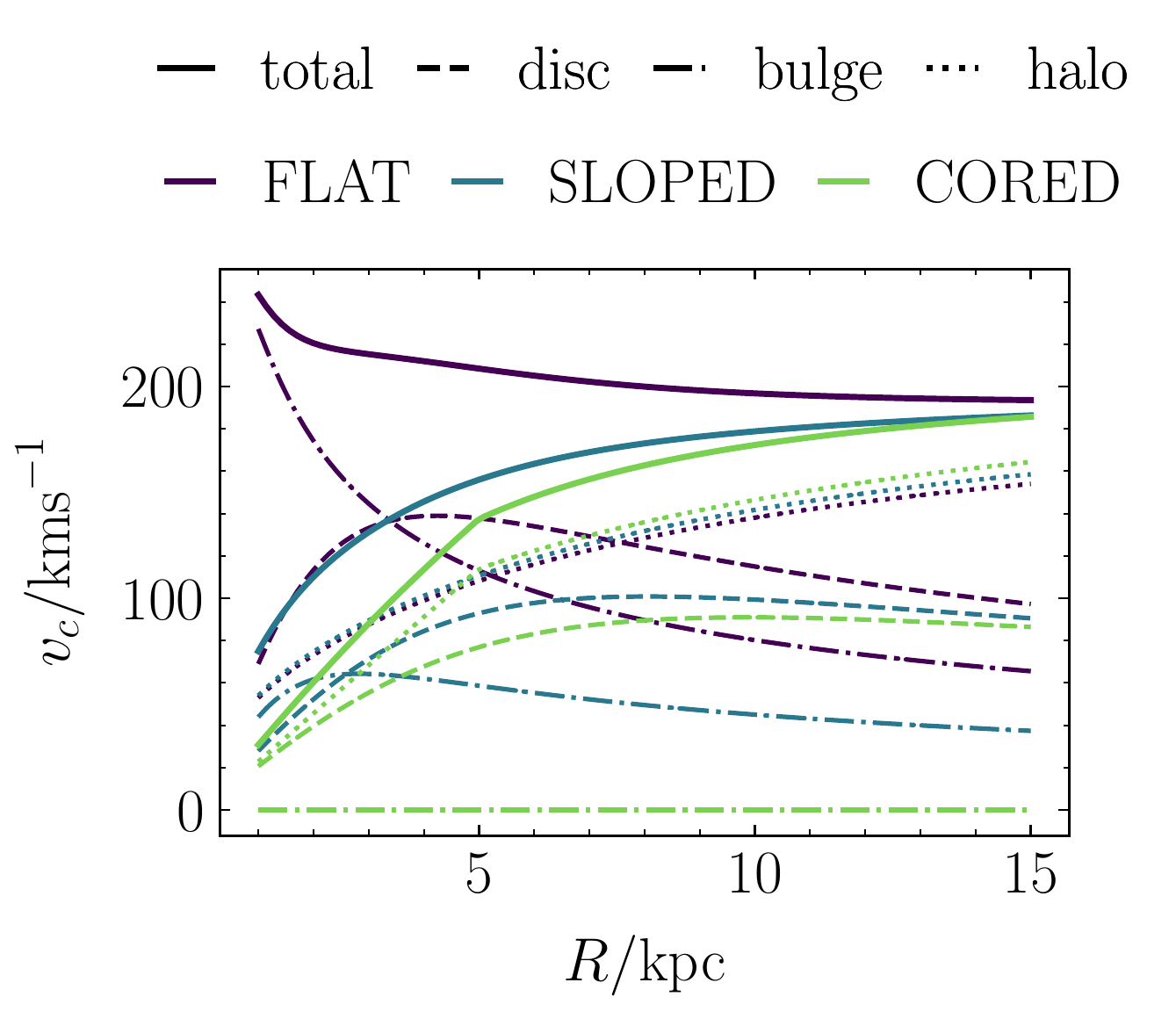}
  \caption{Profiles of the galactic circular velocity for each of the simulated disc galaxies (solid lines). The contributions made by each component of the applied external gravitational potential are illustrated by the dashed lines (disc), dash-dotted lines (bulge) and dotted lines (halo).}
\end{figure}

We analyse the lifetimes of molecular clouds across the three Milky Way-like isolated galaxy simulations of~\cite{Jeffreson20}, shown in Figure~\ref{Fig::morphology}. Here we briefly describe the most important characteristics of our numerical method, and refer the reader to the cited work for a fuller and more detailed explanation.

\begin{table}
\begin{center}
\label{Tab::params}
  \caption{Physical parameters for the disc galaxies modelled in this work, including the masses of each component of the background potential, $M_{\rm h}$ (halo), $M_{\rm b}$ (bulge) and $M_{\rm d}$ (disc). Properties of the gas disc ($M_{\rm gas}$, $R_{\rm gas}$ and $z_{\rm gas}$) are quoted at a simulation time of $\sim 600$~Myr. All masses are given in units of $10^{10} {\rm M}_\odot$, and all length-scales are given in units of kpc. The columns report: (1) Halo mass, (2) bulge mass, (3) disc mass, (4) gas disc mass, (5) gas disc scale-radius, (6) gas disc scale-height.}
  \begin{tabular}{@{}l c c c c c c@{}}
  \hline
   Sim. & $M_{\rm h}$ & $M_{\rm b}$ & $M_{\rm d}$ & $M_{\rm gas}$ & $R_{\rm gas}$ & $z_{\rm gas}$ \\
    & (1) & (2) & (3) & (4) & (5) & (6) \\
  \hline
  \hline
   FLAT & 116 & 1.5 & 3.5 & 0.58 & 7.4 & 0.38 \\
   SLOPED & 130 & 0.5 & 3.5 & 0.59 & 7.7 & 0.28 \\
   CORED & 150 & - & 3.5 & 0.6 & 7.4 & 0.25 \\
  \hline
\end{tabular}
\end{center}
\end{table}

\subsection{Isolated galaxy models} \label{Sec::isolated-galaxies}
The initial conditions for each isolated disc galaxy are generated using {\sc MakeNewDisk}~\citep{Springel2005}, using a three-component external potential consisting of a spherical \cite{Hernquist90} dark matter halo, a~\cite{Miyamoto&Nagai75} stellar disc, and a~\cite{Plummer1911} stellar bulge. The gas disc follows an exponential density profile of the form
\begin{equation}
\label{Eqn::exp-disc}
\rho_{\rm g}(R,z) \frac{M_{\rm g}}{4\pi R_{\rm g}h_{\rm g}} \exp{\Big(-\frac{R}{R_{\rm g}}\Big)} \exp{\Big(-\frac{|z|}{h_{\rm g}}\Big)},
\end{equation}
where $M_{\rm g}$ is the total gas mass and $h_{\rm g}$ is the disc scale-height set by the condition of hydrostatic equilibrium for a mono-atomic gas governed by a polytropic equation of state with a specific heat capacity ratio of $\gamma = 5/3$. The gas disc scale-length is given by $R_{\rm g}$, which is fully-determined by the external potential. We vary the external potential to set three different galactic rotation curves as shown in Figure~\ref{Fig::rotcurve-components}, ensuring that each simulation spans a different set of galactic-dynamical environments. The final gas mass, scale-height and scale-length of each disc are given in Table~\ref{Tab::params}, along with the masses of each component of the external potential.

Each simulation refines adaptively to a target gas cell mass of $900$~M$_\odot$. We avoid artificial fragmentation at scales larger than the Jeans length by ensuring that the disc scale-height and Toomre mass are resolved at all scales~\citep{Nelson06}, and by employing the adaptive gravitational softening scheme in {\sc Arepo}, with a typical value of $1.5$ times the Voronoi cell size and a minimum value of $\sim 3$~pc, corresponding to the spatial resolution in the densest gas at our star formation threshold of $n_{\rm thresh} = 2000$~H/cc.\footnote{As discussed in Section 2.3 of~\cite{Jeffreson20}, we do not impose an artificial non-thermal pressure floor, as this would require us to inflate the Jeans length inside our molecular clouds to unphysically-high values of $\sim 10$~pc, suppressing the unresolved (but physical) gravitational fragmentation required to obtain densities exceeding $n_{\rm thresh}$.} As such, stars are formed only from gas cells that are gravitationally-collapsing (i.e.~whose masses safely exceed the Jeans mass), assuming that the star-forming gas is in approximate thermal equilibrium and has a maximum temperature of $100$~K. Our star formation prescription is chosen to locally reproduce the observed relation of~\cite{Kennicutt98} between the SFR surface density and the gas surface density, following the equation
\begin{equation}
\label{Eqn::starformation}
\frac{\dd \rho_{*,i}}{\dd t} = 
\begin{cases}
   \frac{\epsilon_{\rm ff} \rho_i}{t_{{\rm ff},i}}, \; n_i \geq n_{\rm thresh} \\
   0, \; n_i < n_{\rm thresh}\\
  \end{cases},
\end{equation}
where the local free-fall time within each gas cell is given by $t_{{\rm ff},i} = \sqrt{3\pi/(32G \rho_i)}$ for a mass volume density of $\rho_i$. We use a star formation efficiency per free-fall time of $\epsilon_{\rm ff} = 1$~per~cent, following measurements of the gas depletion time across nearby galaxies~\citep{Leroy17,Krumholz&Tan07,Krumholz18,Utomo18}.

The star particles generated via this prescription each spawn a stellar population drawn stochastically from a~\cite{Chabrier03} initial stellar mass function (IMF) using the Stochastically Lighting Up Galaxies (SLUG) population synthesis model~\citep{daSilva12,daSilva14,Krumholz15}. At each time-step, SLUG provides the ionising luminosity for each star particle, along with the number of supernovae (SN) it has generated and the mass it has ejected, by evolving the stellar populations along Padova solar metallicity tracks~\citep{Fagotto94a,Fagotto94b,VazquezLeitherer05} using {\sc Starburst99}-like spectral synthesis~\citep{Leitherer1999}. Stellar feedback from stellar winds consists of mass ejected from star particles without any accompanying SN events.

In addition to the mass from stellar winds, we include pre-SN photo-ionisation feedback from HII regions, according to the prescription of~\citet{Jeffreson20c}. We inject momentum into the gas cells that share a face with the central `host' cell for each star particle, corresponding to the momentum due to gas and radiation pressure from `blister-type' HII regions, following the analytic description of~\cite{Matzner+02,Krumholz&Matzner09}. The ionisation front radius of the HII region associated with each star particle is calculated, and used to group the star particles via a friends-of-friends prescription, improving the numerical convergence of the feedback model. The resulting Str\"{o}mgren radii are at best marginally-resolved at our mass resolutions, and the gas cells inside these radii are self-consistently heated and held above a temperature floor of $7000$~K, for as long as they continue to receive ionising photons from the star particles. We do not explicitly adjust the chemical state of the heated gas cells, but rely on the chemical network to ionise the gas in accordance with the injected thermal energy. We inject mechanical SN feedback according to the prescription of~\cite{Keller19}, which computes the terminal momentum of the (unresolved) SN remnant according to~\cite{Gentry17}. The energy and momentum injected into each gas cell from all types of stellar feedback is weighted according to the face area shared between the central and receiving gas cells as in~\cite{Hopkins18b}. However, for the HII region feedback we re-weight the momentum along a directed beam of the form
\begin{equation}
\label{Eqn::HII-directed-mom}
\begin{split}
\Delta p_{k,{\rm HII}} &= w_k(\phi_k, A_k) \hat{\mathbf{r}}_{j \rightarrow k} \Delta p_{j, {\rm HII}} \\
w(\phi_k, A_k) &= \frac{A_{j \rightarrow k} f(\phi_k)}{\sum_k A_{j \rightarrow k} f(\phi_k)} \\
f(\phi_k) &= \Big[\log{\Big(\frac{2}{\Theta}\Big)}(1+\Theta^2 - \cos^2{\phi_k})\Big]^{-1},
\end{split}
\end{equation}
where $\Delta p_{j, {\rm HII}}$ is the total momentum delivered to the central cell $j$ that hosts the HII region, $\Delta p_{k, {\rm HII}}$ is the fraction of the momentum injected into the $k$th momentum-receiving cell, $\hat{\mathbf{r}}_{j \rightarrow k}$ defines the axis joining the centroids of the cells $j$ and $k$, and $w_k(\phi_k, A_k)$ is the final weight-factor, dependent on the facing area $A_k$ between the two cells and the angle $\phi_k$ between the axis $\hat{\mathbf{r}}_{j \rightarrow k}$ and the axis of the beam along which the momentum is directed. The direction of the beam is chosen at random from a uniform spherical distribution, and the opening angle is set to a fiducial value of $\Theta=\pi/12$ radians.\footnote{We have also tested a larger opening angle of $\Theta=\pi/6$ radians, and find that this makes no difference to our results.} Our motivation for using beamed feedback is to emulate the `rocket effect', whereby the momentum imparted to the cloud is directional as a result of photoionised gas break-out.

Throughout each simulation, the thermal and chemical state of the simulated gas is determined via the chemical network of~\cite{NelsonLanger97,GloverMacLow07a,GloverMacLow07b}, according to a simplified set of reactions that follow the fractional abundances of ${\rm H}$, ${\rm H}_2$, ${\rm H}^+$, ${\rm C}^+$, ${\rm CO}$, ${\rm O}$ and ${\rm e}^-$, with the abundances of Helium, silicon, carbon and oxygen set to their solar values of $x_{\rm He} = 0.1$, $x_{\rm Si} = 1.5 \times 10^{-5}$, $x_{\rm C} = 1.4 \times 10^{-4}$ and $x_{\rm O} = 3.2 \times 10^{-4}$, respectively. The strength of the interstellar radiation field (ISRF) is set to a value of $1.7$~\cite{Habing68} units according to~\cite{Mathis83}, and a value of $2 \times 10^{-16}$~s$^{-1}$ is used for the cosmic ray ionisation rate~\citep[e.g.][]{Indriolo&McCall12}. A full list of the heating and cooling processes considered in our simulations is given in~\cite{Jeffreson20}, and a detailed account of the chemical network and its coupling to the thermal and dynamical evolution of the gas is given in~\cite{GloverMacLow07a,GloverMacLow07b,Glover10}.

\subsection{Chemical post-processing} \label{Sec::chem-thermal-postproc}
As described in~\cite{Jeffreson20}, we compute the molecular hydrogen abundances of the Voronoi gas cells in our simulations in post-processing, using the {\sc Despotic} model for astrochemistry and radiative-transfer~\citep{Krumholz13a}. Although our run-time chemical network produces a molecular hydrogen fraction, at our mass resolution of $\sim 900~{\rm M}_\odot$, the self-shielding of molecular hydrogen from the UV radiation field cannot be accurately computed on-the-fly. This results in an under-estimation of the molecular mass by a factor of $\sim 2$, requiring us to re-calculate an equilibrium molecular fraction during post-processing. Each gas cell is treated as a one-zone spherical `cloud' with a hydrogen atom number density $n_{\rm H}$, a column density $N_{\rm H}$ and a virial parameter $\alpha_{\rm vir}$. The escape probability formalism is applied to compute the line emission from each cell, coupled self-consistently to the chemical and thermal evolution of the gas. The carbon and oxygen chemistry is followed via the chemical network of~\cite{Gong17}, while the calculation of the temperature includes heating by cosmic rays and the grain photo-electric effect, subject to dust- and self-shielding for each component, line cooling due to ${\rm C}^+$, ${\rm C}$, ${\rm O}$ and ${\rm CO}$, and thermal exchange between dust and gas. The ISRF strength and cosmic ionisation rate are matched to those used to compute the live chemistry during run-time. The entire system of coupled rate equations is converged to a state of chemical and thermal equilibrium for each one-zone model.

Due to considerations of computational cost, we do not perform the above convergence calculation for all gas cells in the simulation, but instead interpolate over a table of pre-calculated models at logarithmically-spaced values of $n_{\rm H}$, $N_{\rm H}$ and $\alpha_{\rm vir}$. For a gas cell with mass density $\rho$, the hydrogen number density is given by
\begin{equation}
n_{\rm H} = \frac{\rho}{\mu m_{\rm H}},
\end{equation}
where $m_{\rm H}$ is the proton mass and $\mu = 1.4$ is the atomic mass per hydrogen nucleus at the standard cosmic composition. The hydrogen column density is then obtained following~\cite{Fujimoto19}, via the local approximation of~\cite{Safranek-Shrader+17}, as
\begin{equation}
N_{\rm H} = \lambda_{\rm J} n_{\rm H},
\end{equation}
where $\lambda_{\rm J} = (\pi c_s^2/G\rho)^{1/2}$ is the Jeans length, and is calculated with an upper limit of $T=40$~K on the gas cell temperature. Finally, the virial parameter is defined according to~\cite{MacLaren1988,BertoldiMcKee1992}, as
\begin{equation}
\alpha_{\rm vir} = \frac{5\sigma_{\rm g}^2}{\pi G \rho L^2},
\end{equation}
where $\sigma_{\rm g}$ is the turbulent velocity dispersion of the gas cell following~\cite{Gensior20}, and $L$ is the smoothing length over which $\sigma_{\rm g}$ is calculated. Using the above three values, we constrain the $^{12}{\rm CO}$ line luminosity $L_{\rm CO}$ for the $1 \rightarrow 0$ transition, from which we obtain the CO-bright molecular hydrogen surface density,\footnote{We note that $\Sigma_{\rm H_2}$ is not the true molecular hydrogen column density in CO-dominated gas cells, but specifically the molecular hydrogen column density that would be inferred by an observer who assumed a fixed CO-to-${\rm H_2}$ conversion factor of $\alpha_{\rm CO} = 4.3~{\rm M}_\odot~({\rm K}~{\rm kms}^{-1} {\rm pc}^{-2})^{-1}$.} according to
\begin{equation}
\begin{split}
\Sigma_{\rm H_2}[{\rm M}_\odot {\rm pc}^{-2}] = &\frac{2.3 \times 10^{-29}[{\rm M}_\odot ({\rm erg} \: {\rm s}^{-1})^{-1}]}{m_{\rm H}[{\rm M}_\odot]} \\
&\times \int^\infty_{-\infty}{dz^\prime \rho_{\rm g}(z^\prime) L_{\rm CO}[{\rm erg} \: {\rm s}^{-1}]}.
\end{split}
\end{equation}
In the above, $\rho_{\rm g}(z)$ is the total gas volume density as a function of distance $z$ from the galactic mid-plane and $\Sigma_{\rm g}$ is the total gas surface density. The mass-to-luminosity conversion factor $\alpha_{\rm CO}=4.3~{\rm M}_\odot~({\rm K}~{\rm kms}^{-1} {\rm pc}^{-2})^{-1}$ of~\cite{Bolatto13} and the line-luminosity conversion factor $5.31 \times 10^{-30} ({\rm K} \: {\rm kms}^{-1} {\rm pc}^2)/({\rm erg} \: {\rm s}^{-1})$ of~\cite{SolomonVandenBout05} for the CO $J=1\rightarrow 0$ transition at redshift $z = 0$ are combined to produce the factor of $2.3 \times 10^{-29} \: ({\rm erg} \: {\rm s}^{-1})^{-1}$. The ratio of integrals represents the two-dimensional density-weighted ray-tracing map of the CO line-luminosity.

\section{Construction of the cloud evolution network} \label{Sec::construct-cloud-history}
In this section, we describe the construction of detailed \textit{cloud evolution networks} from the simulations outlined in Section~\ref{Sec::simulations}. We produce each network at the (two-dimensional) native resolution of $\epsilon = 6$~pc for our simulations, as well as at degraded spatial resolutions of $12$, $18$, $36$, $78$, $198$ and $396$~pc, to examine the time-dependent properties of molecular clouds as a function of their spatial scale. We use a range of simulation times between $600$ and $1000$~Myr, for which the simulated galaxies are in a state of dynamical equilibrium~\citep{Jeffreson20}.

\subsection{Cloud identification} \label{Sec::cloud-ID}
\begin{figure}
 \label{Fig::mass-veldisp-before-prune}
  \includegraphics[width=\linewidth]{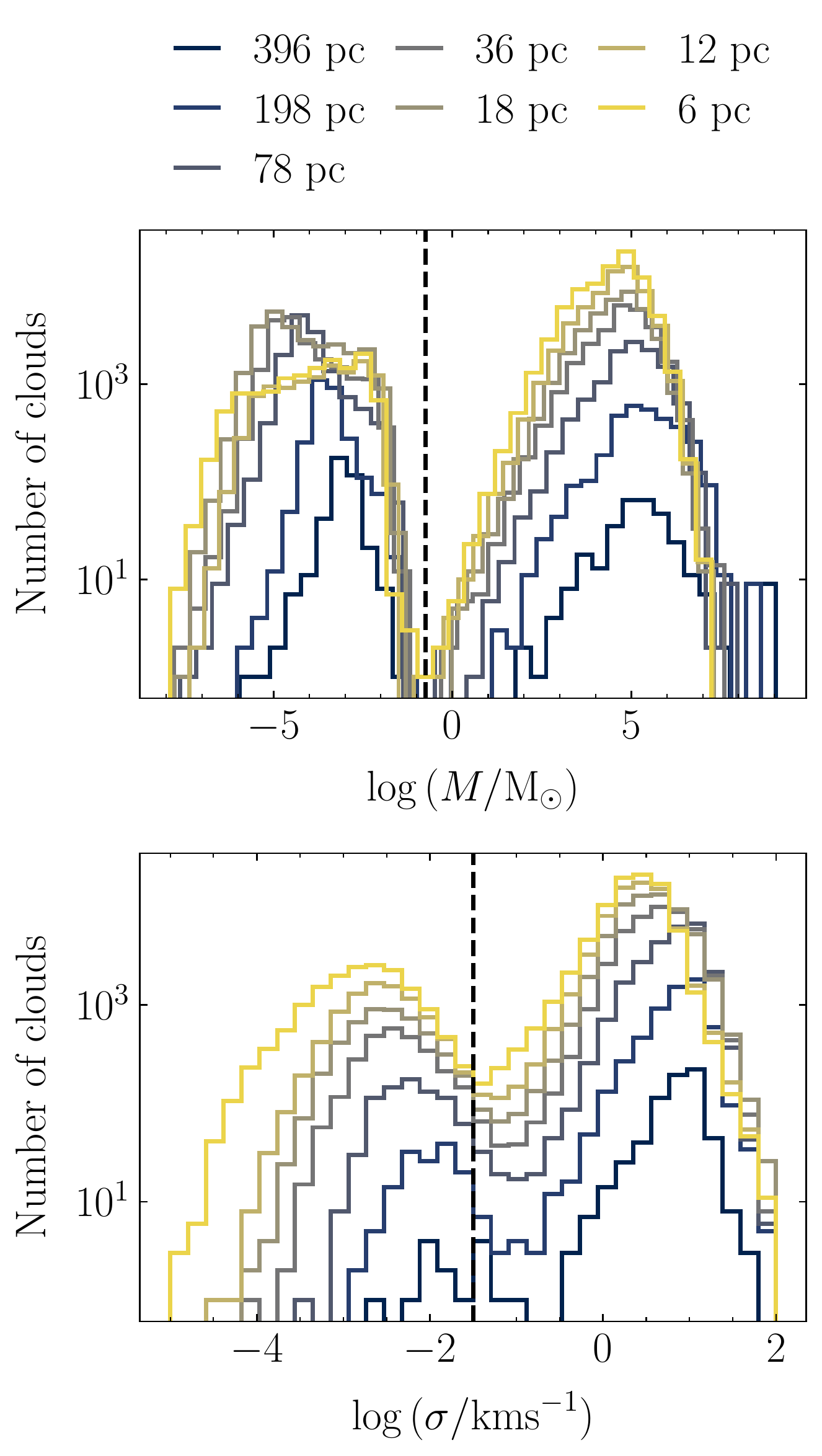}
  \caption{Distributions of cloud masses $M$ and velocity dispersions $\sigma$ in the cloud evolutionary network at each of the different spatial resolutions for cloud identification, before pruning is applied (see Section~\ref{Sec::cloud-pruning}). There exists a large population of low-mass, low-velocity dispersion artefacts (left side of the vertical black dashed lines) that do not conform to observations of the cloud mass spectrum or velocity dispersion distribution. These artefacts are removed from the network before any analysis is performed.}
\end{figure}
\begin{figure*}
 \label{Fig::mass-size-after-prune}
  \includegraphics[width=\linewidth]{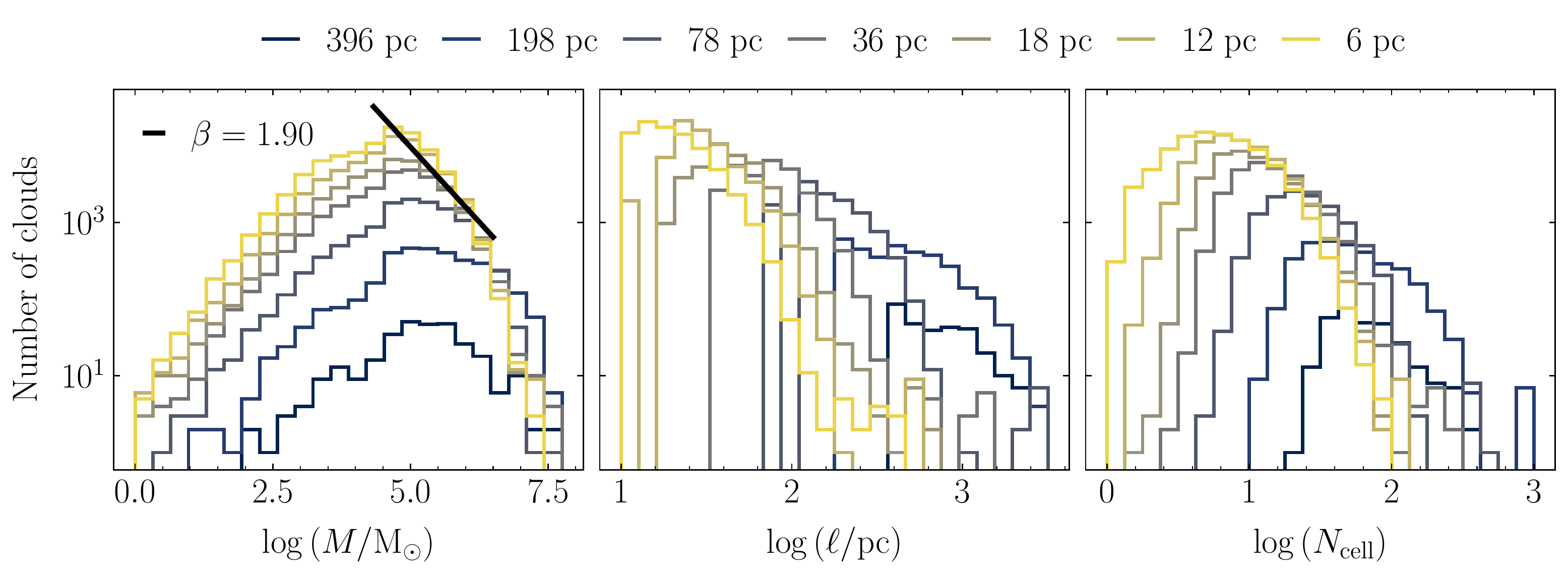}
  \caption{Distributions of masses $M$ (left-hand panel), scales $\ell$ (central panel) and the number of Voronoi cells within each simulated cloud $N_{\rm cell}$ (right-hand panel) in the FLAT cloud evolution network, after the removal of low-mass, low-velocity dispersion artefacts (see Section~\ref{Sec::cloud-pruning}). In the left-hand panel, the solid black line gives the power-law fit to the mass spectrum, with $\beta = 1.90 \pm 0.08$, where $\dd N/\dd M \propto M^{-\beta}$. We see that after pruning, 99.3~per~cent of clouds across all spatial resolutions are resolved by 10 or more Voronoi cells, and have masses that are consistent with the observed distribution, to within the limitations of observational resolution. We also note that the typical cloud size varies systematically with the spatial resolution.}
\end{figure*}
We identify giant molecular clouds at each simulation snapshot using a procedure similar to that described in~\cite{Jeffreson20}. We compute a map of the molecular hydrogen surface density $\Sigma_{\rm H_2}$ at a spatial resolution of $6$~pc, using {\sc Arepo}'s ray-tracing algorithm. For the typical gas cell mass of $\sim 900~M_\odot$, this is equal to the radius of a Voronoi gas cell at the minimum volume density of $n_{\rm H} \ga 30~{\rm cm}^{-3}$ inside molecular clouds, ensuring that each pixel in every cloud contains at least one cell centroid. To obtain the maps at degraded spatial resolutions of $\epsilon = 6$, $12$, $18$, $36$, $78$, $198$ and $396$~pc, we downsample the original at factors of $2$, $3$, $6$, $13$, $33$ and $66$ times, by averaging the value of $\Sigma_{\rm H_2}$ across groups of adjacent pixels. The cloud populations at each degraded resolution are identified from the same simulated interstellar medium as is the cloud population at the native resolution. That is, the lower-resolution clouds correspond to lower-density levels of the same hierarchical interstellar medium.

As described in Section 2.9.1 and in Figure 4 of~\cite{Jeffreson20}, we use the {\sc Astrodendro} package for Python to identify molecular clouds as a set of closed contours at a CO-bright surface density of $\log_{10}{(\Sigma_{\rm H_2}/{\rm M}_\odot {\rm pc}^{-2})} = -3.5$. We use only the `trunk' of the dendrogram, as the cloud sub-structure is not well-resolved at our native resolution. This captures all of the dense CO-dominated gas shielded from the UV radiation field within the {\sc Despotic} model.\footnote{We use this lenient threshold because it corresponds to a natural break in the ${\rm H}_2$ surface density distribution produced by our chemical post-processing: on one side are the cells that contain at least some shielded, CO-dominated gas, and on the other side lie the cells for which the ${\rm H}_2$ and CO exist as uniformly-mixed, unshielded, very low-abundance components. This is an alternative to taking an arbitrary higher-density cut-off, as discussed in~\protect\cite{Jeffreson20}. However, the choice matters very little, because most of the CO and ${\rm H}_2$ reside in cells at much higher density. Increasing the threshold to $10~{\rm M}_\odot~{\rm pc}^{-2}$ affects the total surface area of identified clouds by $<5$~per~cent at the native resolution.} The Voronoi gas cells associated with each cloud are then obtained by applying the {\sc Astrodendro} pixel mask for each cloud to the positions of the gas cell centroids with temperatures $T < 10^4$~K. In contrast to~\cite{Jeffreson20}, at this stage of the analysis we do not impose any requirement on the number of pixels or on the number of Voronoi cells spanned by each cloud. That is, we allow clouds with a diameter of just one pixel, containing one Voronoi cell. This ensures consistency of the cloud identification procedure across maps of varying spatial resolution, which is necessary to compute the scaling relations presented in Sections~\ref{Sec::fractal-structure} and~\ref{Sec::cloud-lifetime} without introducing a spurious bias at the highest resolutions (affecting the smallest scales). However, it produces a population of unphysical artefacts of low mass $M$ and low velocity dispersion $\sigma$, which do not adhere to the observational properties of real molecular clouds, as shown on the left-hand side of the vertical dashed lines in Figure~\ref{Fig::mass-veldisp-before-prune}. Following the construction of the cloud evolution network, we `prune' these artefacts away, according to the physical requirement that our identified clouds reproduce the observed distributions of $M$ and $\sigma$, as described in Section~\ref{Sec::cloud-pruning}. 

\subsection{Tracking clouds over time} \label{Sec::cloud-tracking}
Once we have identified the molecular clouds at every simulation time-step, we track their evolution as a function of time. To identify similar clouds in consecutive snapshots at times $t = t_1$ and $t = t_2 = t_1 + \Delta t$, where $\Delta t = 1$~Myr for our maps, we take the sets of gas cells comprising the clouds identified at $t = t_1$ and calculate their projected positions at $t_2$, according to
\begin{equation} \label{Eqn::cloud-tracking}
\begin{split}
x_2 &= x_1 + v_x \Delta t \\
y_2 &= y_1 + v_y \Delta t. \\
\end{split}
\end{equation}
We then use {\sc Astrodendro} to compute the two-dimensional pixel masks for the closed contours around the time-projected gas cell positions, following the original cloud identification procedure. If any pixel in a time-projected mask overlaps with a pixel in one of the cloud masks at time $t=t_2$, then the clouds are considered indistinguishable at the spatial resolution $\epsilon$ and temporal resolution $\Delta t = 1$~Myr used for cloud identification. The clouds at $t=t_1$ are assigned as the parents of the clouds at $t = t_2$. Via this procedure, each cloud can spawn multiple children (\textit{cloud splits}) or have multiple parents (\textit{cloud mergers}). We connect and store the parents and children of every cloud using the {\sc NetworkX} package for python~\citep{NetworkX}, producing a first version of the cloud evolution network, ready for the pruning procedure described in the following section.

\subsection{Pruning the cloud evolution network} \label{Sec::cloud-pruning}
To obtain the final version of each cloud evolution network, we prune away any nodes that do not correspond to physically-reasonable molecular clouds. These artefacts are produced by regions of faint background CO emission modelled in {\sc Despotic}, which appear as over-densities in the molecular hydrogen surface density and so are picked up by the cloud identification procedure described in Section~\ref{Sec::cloud-ID}, but which contain very little CO-luminous mass.

Figure~\ref{Fig::mass-veldisp-before-prune} demonstrates the bi-modality of the resulting cloud mass and velocity dispersion distributions, which provides a natural choice for the pruning requirement. The observable range of masses $M \ga 10^4~{\rm M}_\odot$ and velocity dispersions $0.5 \la \sigma \la 100~{\rm kms}^{-1}$ extends smoothly\footnote{We note that the pruning threshold for the velocity dispersion includes a small portion of the low-$\sigma$ mode for resolutions $\epsilon \ga 18$~pc, however this accounts for $<0.1$~per cent of the identified molecular clouds and is therefore expected to have a negligible effect on our results.} down to $M \sim 0.2$~M$_\odot$ and $\sigma \sim 0.03$~kms$^{-1}$. Pruning at these cut-offs removes the unphysical artefacts, leaving the spectra presented in Figure~\ref{Fig::mass-size-after-prune} for the cloud mass (left-hand panel), cloud diameter (central panel), and Voronoi cell number $N_{\rm cell}$ (right-hand panel). In the pruned sample, 99~per~cent of clouds across all spatial resolutions, and 74~per~cent at the native resolution, are resolved by 10 or more Voronoi cells (note that the cloud mass presented in Figures~\ref{Fig::mass-veldisp-before-prune} and~\ref{Fig::mass-size-after-prune} is the CO-luminous gas mass, not the Voronoi cell mass: the latter has a median value of $900~{\rm M}_\odot$ and a minimum value of $\sim 60~{\rm M}_\odot$ in our simulations). The mass spectrum (left-hand panel) is consistent with empirical data over the observationally-constrained mass range, with $\beta = 1.90 \pm 0.08$ for the power-law distribution of cloud number with mass, $\dd N/\dd M \propto M^{-\beta}$. Values of $\beta \in [1.6, 1.8]$ are measured consistently in the Milky Way and across other nearby galaxies~\citep{Solomon87,Williams&McKee97,1998A&A...329..249K,Heyer+01,Rosolowsky03,Roman-Duval+10,Freeman17,Miville-Deschenes17,Colombo+19}. Similarly, the upper truncation mass falls at around $M=10^7 {\rm M}_\odot$ at all resolutions, consistent with the observed range of $\sim 3$ to $8 \times 10^6 {\rm M}_\odot$ in The Milky Way~\citep{Colombo+19}, M33~\citep{Rosolowsky03}, M83~\citep{Freeman17}, and across five other nearby galaxies~\citep{Hughes2016}.

For spatial resolutions of $\epsilon < 396$~pc, our pruning requirements allow a significantly smaller value of the lower truncation mass than can be resolved by observations, but given that the slope of the mass spectrum is smooth all the way down to $M=1$~M$_\odot$, we consider these lower-mass clouds to be physical. The final point to note is that the characteristic scale $\ell$ of the identified clouds varies with spatial resolution. As such, clouds at different spatial resolutions correspond to coherent regions of molecular gas at different hierarchical levels within the interstellar medium. Analysis of cloud properties as a function of spatial scale will be a key feature of the following analysis, allowing for the characterisation of hierarchical structure via `scaling relations', and removing the requirement of an arbitrary spatial scale for cloud identification.

\begin{landscape}
\begin{figure}
 \label{Fig::merger-tree}
  \includegraphics[width=\linewidth]{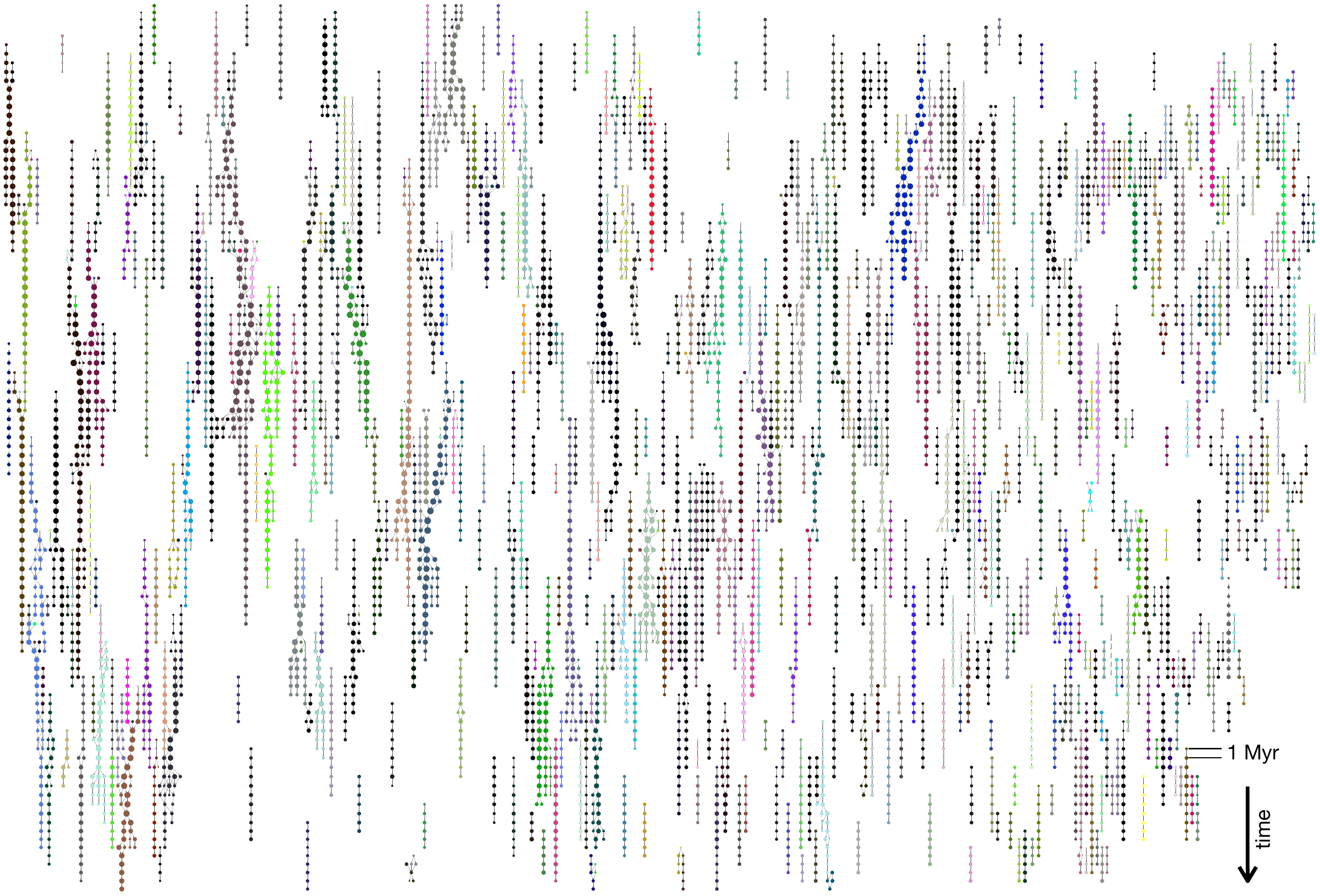}
  \caption{Section of the giant molecular cloud evolution network for the FLAT simulation, spanning the range of simulation times $t \in [600, 700]$~Myr and the range of galactocentric radii $R \in [7.75,8.25]$~kpc (approximately at the solar radius for a Milky Way-like galaxy). The black arrow in the bottom right-hand corner indicates the arrow of time, and each pair of nodes is separated by a time-step of $1$~Myr. The network has been constructed according to the procedures described in Section~\ref{Sec::construct-cloud-history}. The sizes of the nodes scale as the logarithm of cloud mass, and the colours of the nodes follow the evolution of the most massive cloud involved in each split and merger.}
\end{figure}
\end{landscape}

A $100$-Myr section of the final cloud evolution network for the FLAT simulation at the native resolution of $\epsilon = 6$~pc is shown in Figure~\ref{Fig::merger-tree} for galactocentric radii between $7.75$ and $8.25$~kpc (close to the solar radius for a Milky Way-like galaxy). Only clouds that remain inside this annulus for their entire lifetimes are considered. The arrow of time points from the top to the bottom of the network. Each node represents a molecular cloud identified at a single simulation time, as described in Section~\ref{Sec::cloud-ID}. The nodes are separated by a time interval of $\Delta t = 1~{\rm Myr}$, which defines the temporal resolution of the network.

\section{The cloud merger rate} \label{Sec::fractal-structure}
The fractal, self-similar structure of the interstellar medium has been observed in maps probing a large dynamical range in spatial scale and gas density, and spanning a wide variety of different galactic environments~\citep{1985prpl.conf..201S,1987ApJS...65...13B,1990ASSL..162..151S,1990ApJ...355..536L,1991ApJ...378..186F,1991IAUS..147...11B,1996ApJ...471..816E,2009A&A...507..355F}. This spatial distribution of gas is shown to be reproducible via compressible supersonic turbulence in numerical simulations~\citep[e.g.][]{1994ApJ...423..681V,1998PhRvE..58.4501P,1998ApJ...508L..99S,2001ApJ...546..980O,2009ApJ...692..364F,2010A&A...512A..81F}. In this section, we show that the turbulent self-similarity of the interstellar medium also has important consequences for the time-evolution of the giant molecular clouds in our simulations, setting the rate of cloud mergers over scales from $10$~pc to $1$~kpc.

\subsection{Scaling relation of the cloud merger rate} \label{Sec::scaling-rlns}
\begin{figure}
 \label{Fig::no-mergers-per-cloud}
  \includegraphics[width=\linewidth]{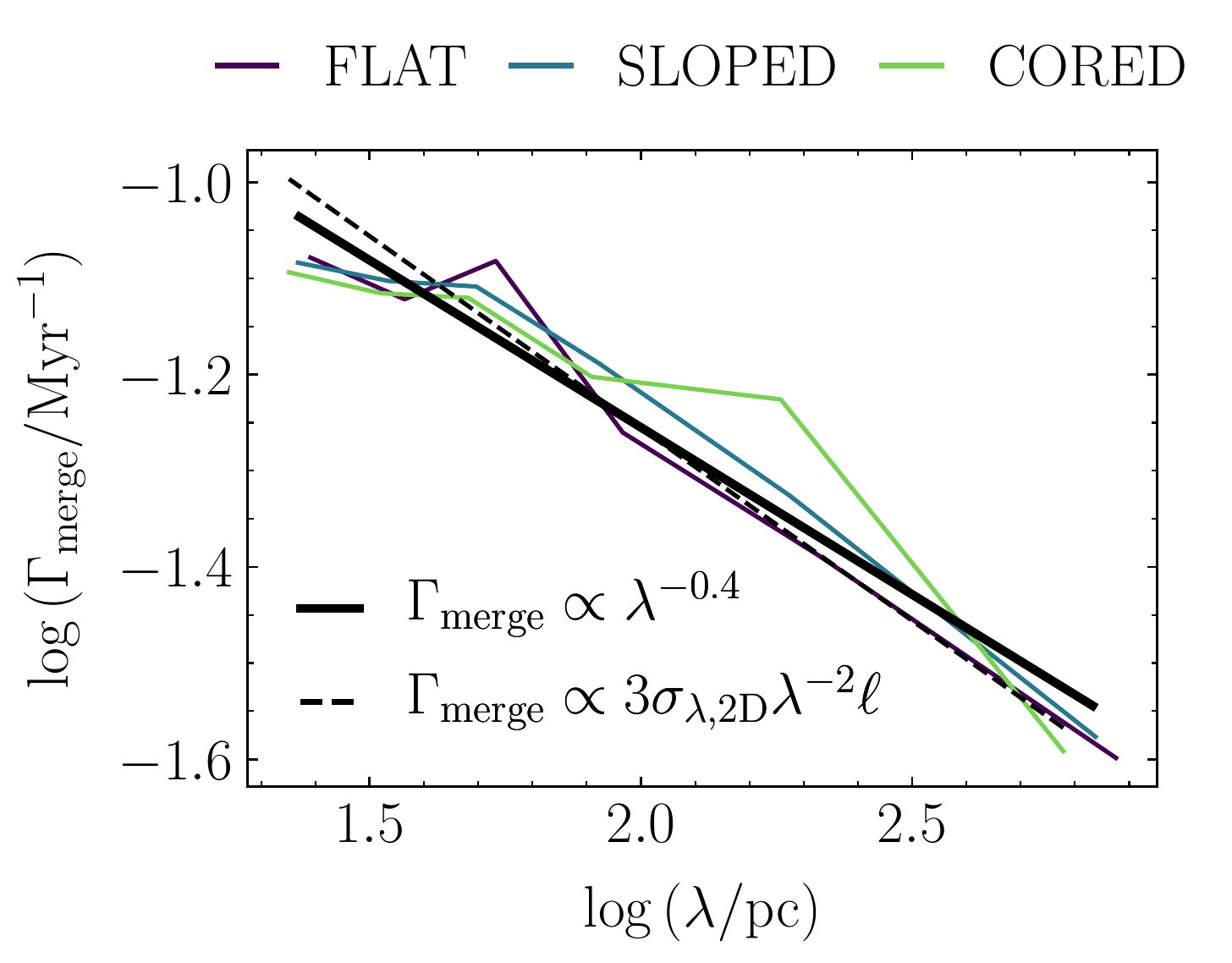}
  \caption{Rate of cloud mergers $\Gamma_{\rm merge}$ (solid lines) as a function of the median cloud separation length $\lambda$ for each simulation. Each data-point is calculated for the entire cloud evolution network at a given resolution, spanning the whole galactic disc across the range of simulation times from $600$~Myr to $1$~Gyr. The solid black line gives the power-law fit to the combined data and the dashed black line gives the best-fit prediction of the collision rate in terms of the crossing time between cloud centroids (see Section~\ref{Sec::scaling-rlns} and Equation~\ref{Eqn::merge-rate}).}
\end{figure}
\begin{figure}
 \label{Fig::size-vs-sep}
  \includegraphics[width=\linewidth]{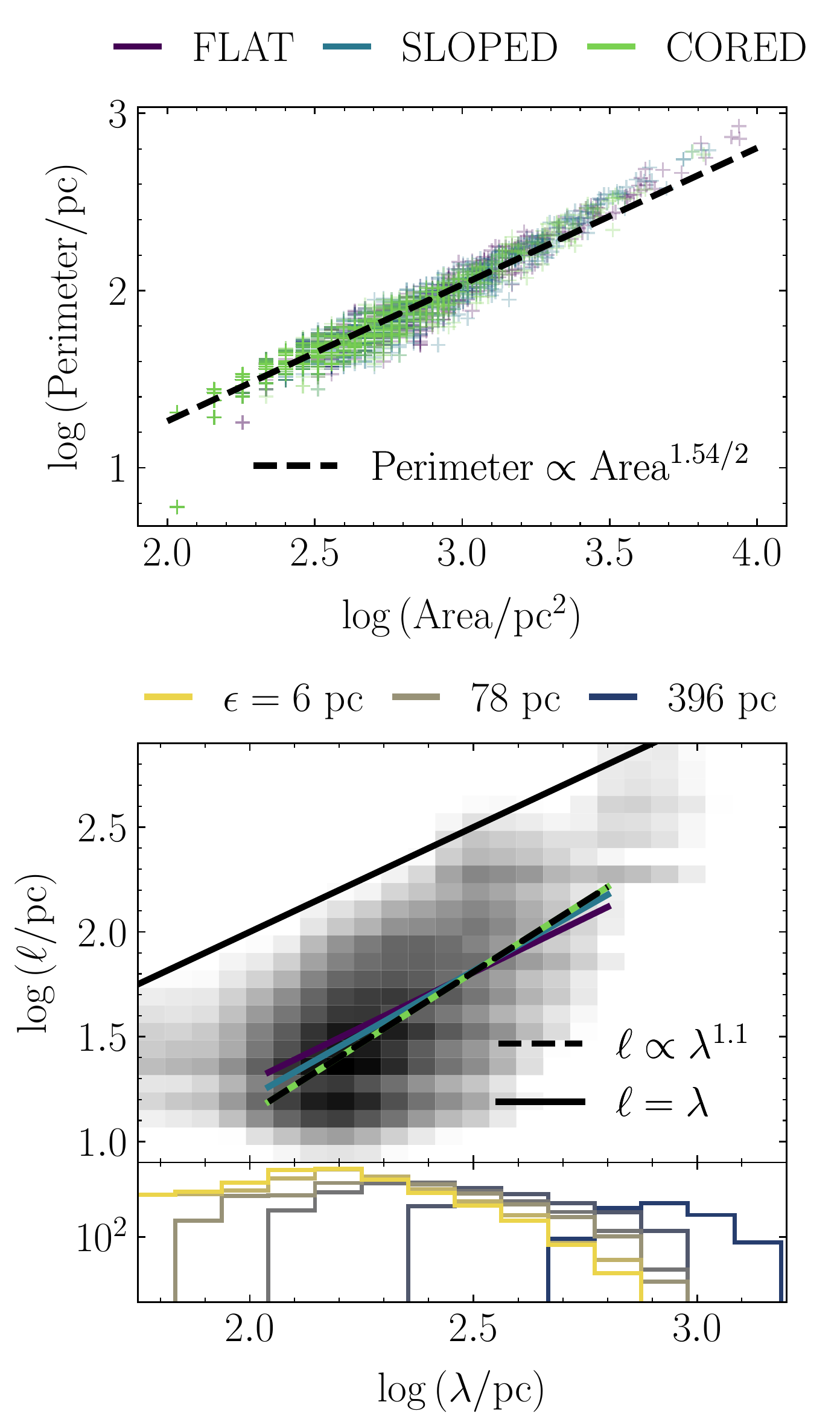}
  \caption{\textit{Top:} Cloud perimeter as a function of cloud area for the combined molecular cloud population across all three simulated disc galaxies, at the native map resolution of $\epsilon = 6$~pc. The black dashed line gives the best-fit power-law to the sample, which yields a fractal dimension of $D = 1.54$. \textit{Bottom:} Cloud size $\ell$ as a function of the cloud separation length $\lambda$. The grey-shaded histogram corresponds to the combined molecular cloud population across all three simulated disc galaxies and across all map resolutions $\epsilon$, sampled at $50$-Myr intervals across the time-span of each cloud evolution network, between simulation times of $600$ and $1000$~Myr. It is displayed on a logarithmic scale with a lower bound of 500 clouds per pixel. The black dashed line shows the best-fit power-law to the combined data. The lower extension panel shows the number of clouds at each separation length that are accounted for by the maps at each spatial resolution, from $\epsilon = 6$~pc up to $\epsilon = 396$~pc.}
\end{figure}
\begin{figure}
 \label{Fig::size-linewidth}
  \includegraphics[width=\linewidth]{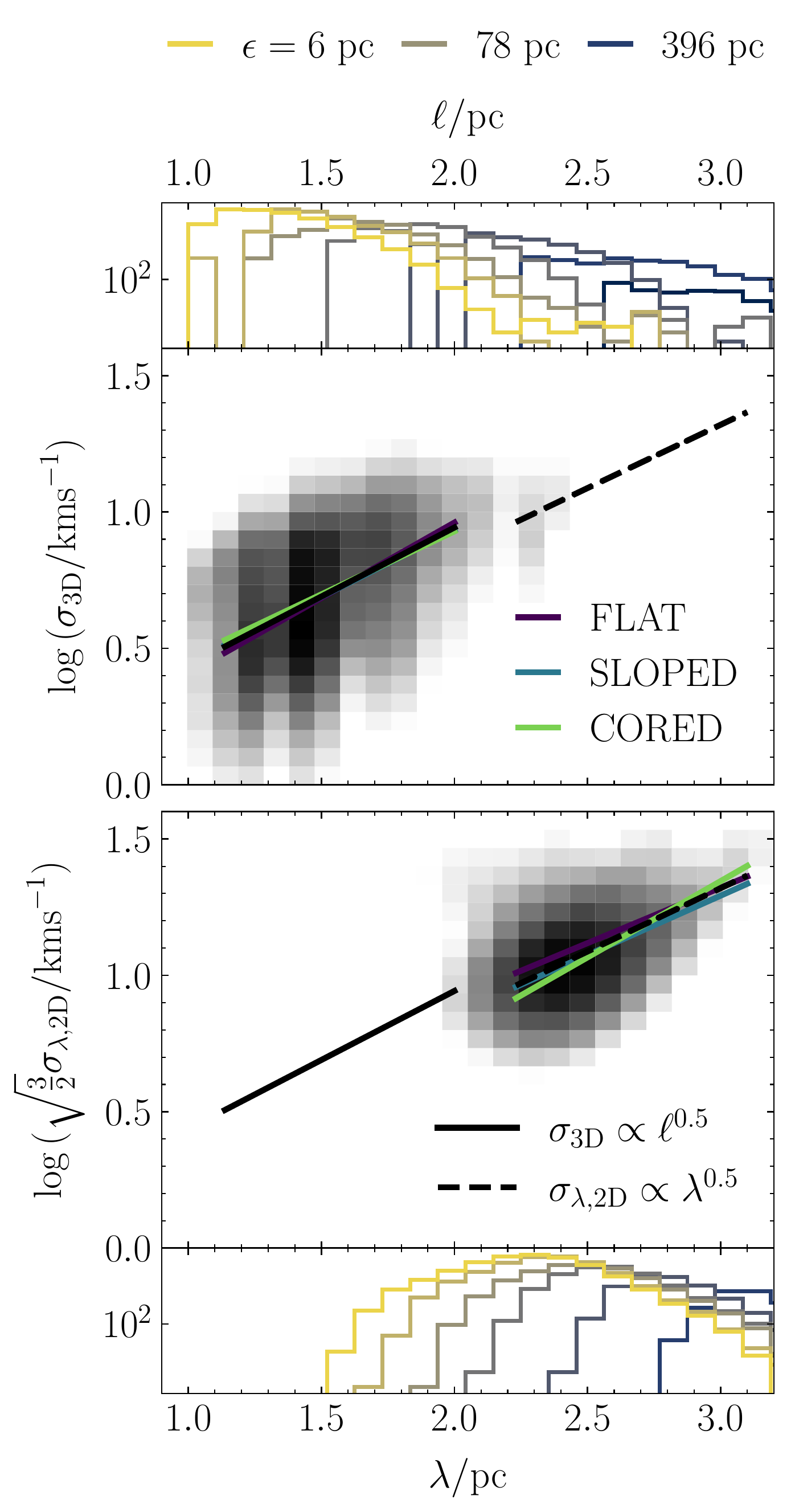}
  \caption{\textit{Top:} Internal three-dimensional velocity dispersion $\sigma_{\rm 3D}$ as a function of the typical cloud scale $\ell$ (size-linewidth relation). \textit{Bottom:} Two-dimensional velocity dispersion of the cloud centroids $\sigma_{\lambda, {\rm 2D}}$ as a function of their typical separation length $\lambda$, scaled by a factor $\sqrt{3/2}$ to enable direct comparison with $\sigma_{\rm 3D}$. The grey-shaded histogram corresponds to the combined molecular cloud population across all three simulated disc galaxies and across all map resolutions $\epsilon$, sampled at $50$-Myr intervals across the time-span of each cloud evolution network, between simulation times of $600$ and $1000$~Myr. It is displayed on a logarithmic scale with a lower bound of 500 clouds per pixel. The solid and dashed black lines, for three and two dimensions respectively, show the best-fit power-laws to the combined data. Both fits are shown on both panels, for reference. The upper and lower extension panels show the number of clouds at each cloud scale $\ell$ and separation length $\lambda$, respectively, that are accounted for by the maps at each spatial resolution, from $\epsilon = 6$~pc up to $\epsilon = 396$~pc.}
\end{figure}

In Figure~\ref{Fig::no-mergers-per-cloud} we show the number of cloud mergers $\Gamma_{\rm merge}$ per unit time in each of our simulations as a function of the median cloud spatial separation $\lambda$. Each data point is calculated for the entire cloud evolution network at each resolution, spanning the entire disc for the set of simulation times ranging from $600$~Myr to $1$~Gyr. The spatial separation at the position of each cloud is calculated across a group of its 100 nearest neighbours, as
\begin{equation} \label{Eqn::lambda}
\lambda = \frac{\lambda_{\rm 100}}{10},
\end{equation}
where $\lambda_{\rm 100}$ is the distance to the furthest nearest neighbour. The number of mergers per unit time is defined as
\begin{equation}
\Gamma_{\rm merge} = \frac{\sum_{\theta=2}^\infty{\theta N_{{\rm merge}, \theta}}}{N_{\rm nodes} \Delta t},
\end{equation}
where $N_{\rm merge, \theta}$ is the total number of merge-nodes in the network involving $\theta$ clouds, $N_{\rm nodes}$ is the total number of nodes in the network, and $\Delta t = 1$~Myr is the temporal separation between nodes. We find that $\theta=2$ in 80~per~cent of cases at the native resolution, with a maximum value of $\theta=8$. The best-fit power-law to the scaling relation is given by the bold black line, with a form of $\Gamma_{\rm merge} \propto \lambda^{-0.4}$. At $\lambda \sim 20$~pc, clouds enter mergers around once in every $10$~Myr; at large $\lambda \sim 500$~pc, the rate drops to once in every $30$~Myr. We can understand the scaling relation of the cloud merger rate by considering the size $\ell$ of a cloud as its `collision cross-section', onto which other clouds impinge. This gives a two-dimensional version of the familiar kinetic collision rate
\begin{equation} \label{Eqn::merge-rate}
\Gamma_{\rm merge} = F \sigma_{\lambda, {\rm 2D}} \lambda^{-2} \ell,
\end{equation}
where $\sigma_{\lambda, {\rm 2D}}$ is the two-dimensional velocity dispersion of the cloud centroids within the galactic mid-plane, and $F$ is a geometric factor accounting for the elongation and orientation of the clouds. In a self-similar interstellar medium the cloud size scales with separation as $\ell \propto \lambda$, so that the merger time-scale is proportional to the crossing time between clouds,
\begin{equation}
\Gamma_{\rm merge} \propto \frac{\sigma_{\lambda, {\rm 2D}}}{\lambda}.
\end{equation}
The dashed black line in Figure~\ref{Fig::no-mergers-per-cloud} gives the merger rate predicted by Equation (\ref{Eqn::merge-rate}) when we substitute the following power-law fits to our simulated cloud population:
\begin{equation} \label{Eqn::fits}
\begin{split}
\ell/{\rm pc} &= 0.12^{+0.06}_{-0.04} (\lambda/{\rm pc})^{1.1 \pm 0.1} \\
\sigma_{\lambda, {\rm 2D}}/{\rm km~s}^{-1} &= 0.71^{+0.17}_{-0.13} (\lambda/{\rm pc})^{0.48 \pm 0.05},
\end{split}
\end{equation}
along with the best-fit geometric factor $F \sim 3$. That is, the cloud merger rate is well-described by the frequency of interactions between molecular clouds in a spatially self-similar interstellar medium ($\ell \propto \lambda$), with random centroid velocities induced by supersonic, compressible turbulence ($\sigma_{\lambda, {\rm 2D}} \propto \lambda^{0.5}$). In the following sub-sections, we describe in more detail each of the scaling relations in Equation (\ref{Eqn::fits}), and evaluate the influence of cloud mergers on the physical properties of the interacting clouds.

\subsubsection{Cloud size vs. cloud separation}
In the lower panel of Figure~\ref{Fig::size-vs-sep}, we show that the relationship of cloud size to cloud spatial separation for our simulations is $\ell \propto \lambda^{1.1}$, not quite the proportionality of $\ell \propto \lambda$ expected in the case of perfect self-similarity. This could be due to our method of calculating the cloud size $\ell$, for which we have used the pixel-by-pixel area of the cloud's footprint on the galactic mid-plane as $\ell = \sqrt{A}$. We show in the top panel of Figure~\ref{Fig::size-vs-sep} that this assumption of approximately-circular clouds with smooth perimeters is not correct: the fractal dimension of the clouds, computed at our native resolution of $6$~pc, is $D = 1.54$, such that the cloud perimeters $P$ scale with their areas as
\begin{equation} \label{Eqn::fractal-scaling}
P \propto A^{D/2}; \; D=1.54.
\end{equation}
This is significantly more complex than the circular case of $D=1$, so the difference relative to the self-similar scaling relation may be due to an over-estimate of the cloud area that worsens at lower spatial resolutions, as the number of pixels characterising each cloud becomes smaller. The deviation of our ISM from perfect self-similarity could also be due to the preferred observable scales imposed by the CO chemistry in our simulations, at low gas densities.

\begin{figure*}
 \label{Fig::converging-flows}
  \includegraphics[width=\linewidth]{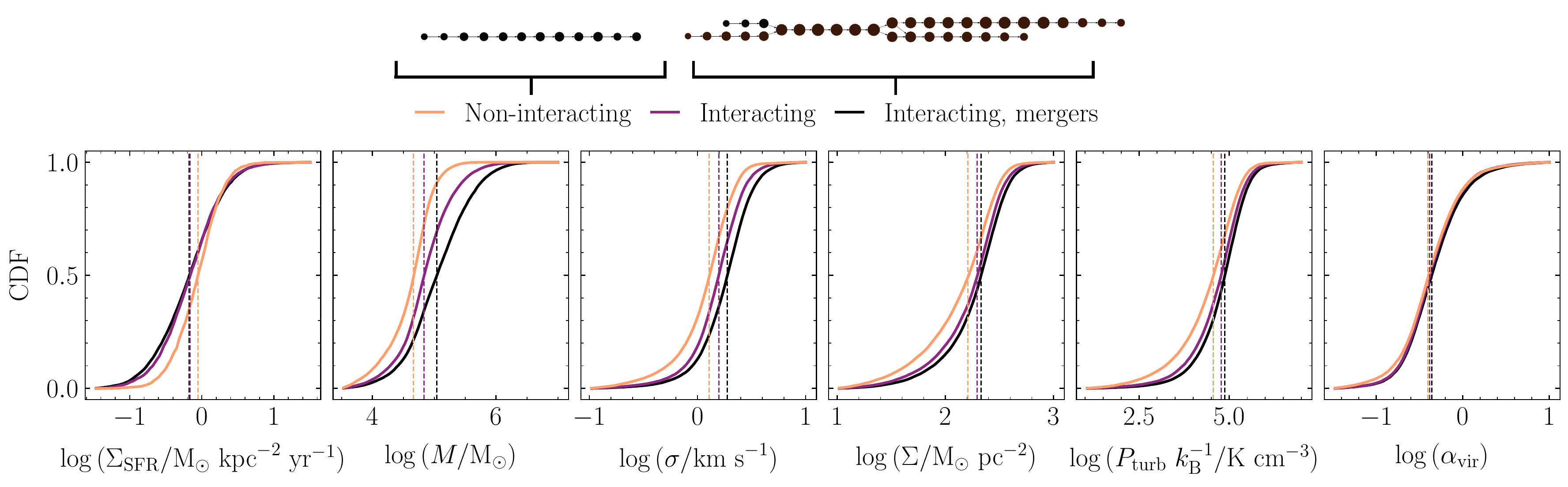}
  \caption{Comparison of the physical properties of segments of cloud evolution for clouds in the FLAT simulation at the native resolution of $\epsilon = 6$~pc that (i) have just undergone a merger (`interacting, mergers', black lines), (ii) undergo a merger or a split at some point in their lives (`interacting', purple lines), or (iii) never undergo a merger or split (`non-interacting', orange lines). A random sample of time-directed evolutionary segments is taken from the cloud evolution network in each case, according to the procedure described in Section~\ref{Sec::converging-flows}. We see that the physical differences between clouds in each case are very small or negligible.}
\end{figure*}

\subsubsection{Cloud centroid velocity dispersion vs. cloud separation}
In the lower panel of Figure~\ref{Fig::size-linewidth}, we show the second scaling relation required to compute $\Gamma_{\rm merge}$ via the cloud collision cross-section of Equation (\ref{Eqn::merge-rate}): the relation of the two-dimensional cloud centroid velocity dispersion $\sigma_{\lambda, {\rm 2D}}$ to the cloud separation length $\lambda$. The value of $\sigma_{\lambda, {\rm 2D}}$ for each cloud is measured across the same $100$ nearest neighbours used to calculate $\lambda$, such that
\begin{equation} \label{Eqn::sigma-lambda}
\sigma_{\lambda, {\rm 2D}} = \sqrt{\langle|v_x - \langle v_x \rangle_{100}|^2 + |v_y - \langle v_y \rangle_{100}|^2\rangle_{100}},
\end{equation}
where $\langle ... \rangle_{100}$ denotes an average over all $100$ neighbours, and $\{v_x, v_y\}$ are the $x$- and $y$-components of their centroid velocities. We retrieve the same scaling relation as is observed for the three-dimensional internal cloud velocity dispersion $\sigma_{\rm 3D}$ with the cloud size, which is shown for our simulations in the upper panel. That is,
\begin{equation} \label{Eqn::obs-scaling-rln}
\begin{split}
\sigma_{\lambda, {\rm 2D}} &\propto \lambda^{0.5} \\
\sigma_{\rm 3D} &\propto \ell^{0.5},
\end{split}
\end{equation}
with a vertical offset of $\sim 0.05$~dex that may be explained by the anisotropy of the velocity field on scales close to the gas disc scale-height. In the above, the three-dimensional internal velocity dispersion of each cloud is defined as
\begin{equation} \label{Eqn::sigma}
\sigma_{\rm 3D} = \sqrt{\langle |\mathbf{v}_i-\langle \mathbf{v}_i \rangle_{i, {\rm H_2}}|^2 \rangle_{i, {\rm H_2}}},
\end{equation}
where $\{\mathbf{v}_i\}$ are the velocities of the gas cells within each cloud, and $\langle ... \rangle_{i, {\rm H_2}}$ denotes the molecular gas mass-weighted average over these cells. Both scalings are consistent with the self-similar distribution of velocity dispersions induced by compressible, supersonic turbulence~\citep[e.g.][]{1995MNRAS.277..377P,2007ApJ...665..416K,McKee&Ostriker2007,2013ApJ...763...51F}, as observed in nearby Galactic molecular clouds~\citep[e.g.][]{Ossenkopf2002,2004ApJ...615L..45H,2011ApJ...740..120R}.

Given the self-similar structure of the turbulent interstellar medium, at first glance it is not too surprising that the motions of the molecular cloud centroids obey the same scaling relation as their internal velocity dispersions. After all, the centroids of distinct clouds at high resolution are simply the turbulent sub-structure of a larger cloud at low resolution. What is surprising, however, is that the form of the power-law continues above the thin-disc scale-height, which is $h_{\rm g} \sim 100$~pc in our simulations~\citep{Jeffreson20}, and which places an upper limit on the vertical extent of the clouds. This implies that turbulent fragmentation on scales $\ge 1$~kpc in the galactic mid-plane proceeds independently of fragmentation perpendicular to the galactic mid-plane, consistent with the idea that the fractal spatial structure of the interstellar medium extends up to the scales of galactic spiral arms~\citep{2000ApJ...530..277E,2003ApJ...590..271E,2003ApJ...593..333E}.

\subsection{The physical impact of cloud mergers} \label{Sec::converging-flows}
The impact of cloud mergers on the turbulent and star-forming properties of the interacting clouds is a direct indicator of their role in setting the cloud lifecycle and the galactic star formation rate. We would like to determine whether mergers significantly alter the demographics of the cloud population, or whether the clouds are simply `nudging' each other~\citep[see][]{Dobbs15}.

\begin{figure}
 \label{Fig::cloud-lifetimes}
  \includegraphics[width=\linewidth]{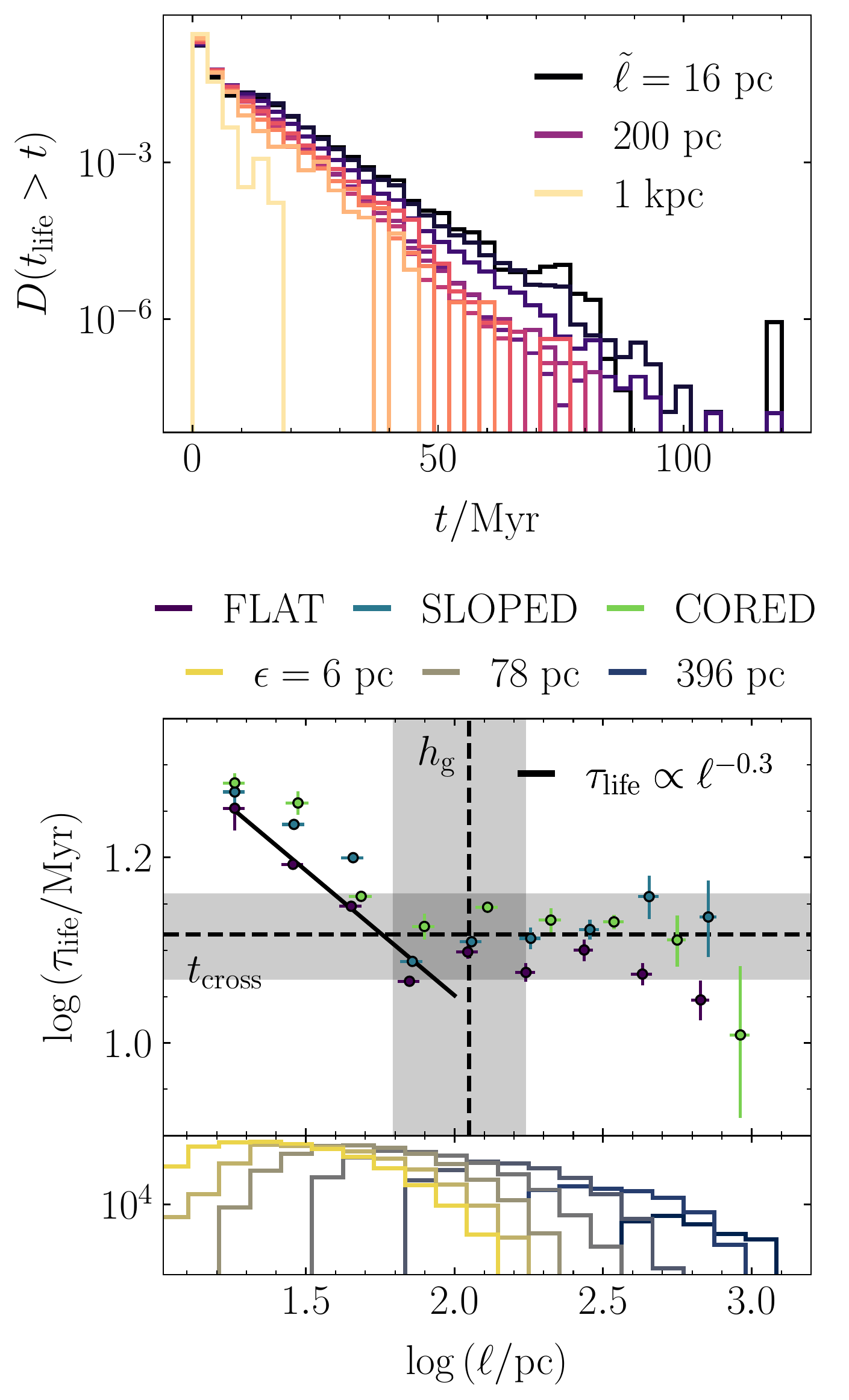}
  \caption{\textit{Top:} Cumulative distribution of trajectory lifetimes $t_{\rm life}$ in ten different bins of cloud size, where smaller sizes correspond to darker colours. The median values $\tilde{\ell}$ of the cloud size in three bins are given by the legend. The exponential form of each distribution is expected for a population of clouds obeying the rate equation (\ref{Eqn::rate-equation}). \textit{Bottom:} Characteristic cloud lifetime $\tau_{\rm life}$ as a function of the cloud size $\ell$ for each simulated galaxy, obtained from the exponential distributions in the top panel by fitting a function $\exp{(-t/\tau_{\rm life})}$, according to Equation~(\ref{Eqn::trajectory-dstbn}). The average values of the gas disc scale-height $h_{\rm g}$ and turbulent crossing time $t_{\rm cross}$ across simulation time, galactocentric radius and galactic azimuthal angle are given by the vertical and horizontal dashed lines, respectively. The corresponding standard deviations are given by the grey-shaded regions. We see that the cloud lifetime obeys a power-law scaling relation $\tau_{\rm life} \propto \ell^{-0.3}$ (solid black line) below the disc scale-height, and converges to the gas disc crossing time above it. The lower extension panel shows the number of clouds at each separation length that are accounted for by the maps at each spatial resolution, from $\epsilon = 6$~pc up to $\epsilon = 396$~pc.}
\end{figure}

To determine the role of mergers in setting the distribution of cloud physical properties, we compare three samples of evolutionary segments from the FLAT cloud evolution network at our native resolution of $\epsilon=6~{\rm pc}$, corresponding to the three lines in each panel of Figure~\ref{Fig::converging-flows}. The samples are defined as follows:
\begin{enumerate}
	\item Black line: Segments that begin at a merger and end at the next merger/split, sampled at random from the population of mergers.
	\item Purple line: Segments sampled at random from components of the network that contain mergers/splits (right-hand schematic in Figure~\ref{Fig::converging-flows}).
	\item Orange line: Segments sampled at random from components of the network that contain no interactions at all (left-hand schematic in Figure~\ref{Fig::converging-flows}).
\end{enumerate}
The lengths of the segments of type (i) determine the lengths sampled for types (ii) and (iii), so that the distribution of segment lengths is identical in all cases. By comparing the physical properties of the three samples, we can answer the following questions:
\begin{itemize}
	\item (i) vs.~(ii): Do mergers affect the physical properties and evolutionary sequences of the merging clouds?
	\item (ii) vs.~(iii): Are the physical properties of interacting clouds different from those of non-interacting clouds?
\end{itemize}
Comparison of the black (i) and purple (ii) lines in Figure~\ref{Fig::converging-flows} demonstrates that cloud mergers cause only a very small change to the physical properties of the clouds in our sample. The distributions of the star formation rate $\Sigma_{\rm SFR}$ per unit area of the galactic mid-plane, the surface density $\Sigma$, the turbulent pressure $P_{\rm turb}$, and the virial parameter $\alpha_{\rm vir}$ are close to identical. The median cloud mass $M$ is slightly elevated for clouds that have recently undergone mergers, which is expected given that the merged cloud is a combination of multiple parents. The median cloud velocity dispersion $\sigma$ is also slightly elevated, which could either be attributed to compression of material at the cloud-cloud interface, or to the increase in cloud mass and size, leading to a higher degree of internal turbulence. The fact that the cloud surface density $\Sigma$ shows no corresponding increase suggests that the latter explanation is more likely.

Comparison of the purple (ii) and orange (iii) lines in Figure~\ref{Fig::converging-flows} demonstrates that the population of interacting clouds differs systematically from the population of clouds that do not interact (although again by only a small amount). On average, interacting clouds have higher masses and surface densities, which in turn leads to a higher degree of turbulence, and so to higher velocity dispersions and turbulent pressures. This depresses the star formation rate surface density relative to that of non-interacting clouds. A detailed analysis of the demographics of these two populations is beyond the scope of the current paper, and so is relegated to future work. We note simply that larger and more massive clouds have larger collision cross-sections, leading to a higher rate of mergers via Equation (\ref{Eqn::merge-rate}). They are also more likely to undergo splitting events, which may re-merge at a later time.

\section{The molecular cloud lifetime} \label{Sec::cloud-lifetime}
We have shown that the fractal spatial structure of the molecular interstellar medium in our simulations sets a merger rate $\Gamma_{\rm merge} \propto \lambda^{-0.4}$ for giant molecular clouds of separation length $\lambda$. Although at our numerical resolution, mergers do not have a large impact on the internal turbulent and star-forming properties of the molecular gas, their occurrence is frequent: almost 80~per~cent of clouds at scales $\ell \sim 10$~pc and separations $\lambda \sim 100$~pc experience a merger during their lifetime. In the following sub-sections, we describe a method for computing the molecular cloud lifetime that takes into account the frequent mergers and splits within the cloud evolutionary network. We calculate this cloud lifetime as a function of spatial scale, and examine its dependence on the large-scale galactic environment.

\subsection{Walking through the cloud evolution network} \label{Sec::random-walk}
We require an approach that describes the distribution of temporal lengths for time-directed trajectories through the cloud evolution network, while accounting for cloud interactions via the following two requirements:
\begin{enumerate}
 \item \textit{Cloud uniqueness:} Each edge connecting two nodes (arrows in Figure~\ref{Fig::merger-tree}) in the network represents a time-step in the evolution of a single cloud, and so can contribute to just one cloud lifetime. Edges must not be double-counted when calculating cloud lifetimes.
 \item \textit{Cloud number conservation:} Each cloud (unique trajectory in Figure~\ref{Fig::merger-tree}) can be formed and destroyed only once, so the number of cloud lifetimes retrieved from the entire network must be equal to the number of cloud formation events and cloud destruction events.
\end{enumerate}

In addition, we avoid making arbitrary choices between cloud-evolutionary paths as they pass through mergers and splits.\footnote{We present here the most basic form of the algorithm, with no assumptions about what constitutes the destruction of a cloud, other than that a node is removed from the cloud evolution network from one time-step to the next. In Section~\ref{Sec::discussion}, we discuss how the algorithm could be altered to distinguish between cloud mergers and cloud accretion.} At a merger involving two clouds $A$ and $B$, there are two mutually-exclusive outcomes: (1) $A$ continues to evolve while $B$ is considered to have been destroyed, and (2) $A$ is destroyed and $B$ continues to evolve. The method for satisfying (i) and (ii) while also sampling from the set of all unique time-directed trajectories through the network is the Monte Carlo (MC) walk described in Appendix~\ref{App::cloud-lifetimes}. For each MC iteration, a number of walkers $N_f$ are initialised at every \textit{formation node} in the network, where \textit{formation nodes} are defined by a net increase in the number of clouds, $N_f > 0$. Each walker steps along the edges between nodes, counting the number of time-steps it takes, until it reaches an \textit{interaction node} (which may also be the formation node itself) with multiple parents or children. A random number from the uniform distribution $U(0,1)$ is assigned to all such interaction nodes for a given MC iteration, and this random number is used to choose between possible subsequent trajectories for the walker, including the possibility of cloud destruction. Upon destruction of a cloud, the walker is terminated, and returns a lifetime $t_{\rm life}$ for the trajectory. Via this algorithm, each edge joining pairs of nodes in the cloud evolution network is visited by a walker exactly once in each MC iteration. We perform 70 such iterations to reach convergence of the characteristic molecular cloud lifetime $\tau_{\rm life}$ for the cloud population of an entire galaxy. This forms the cloud sample analysed in the remainder of this section.

\begin{figure}
 \label{Fig::collapse-profiles}
  \includegraphics[width=\linewidth]{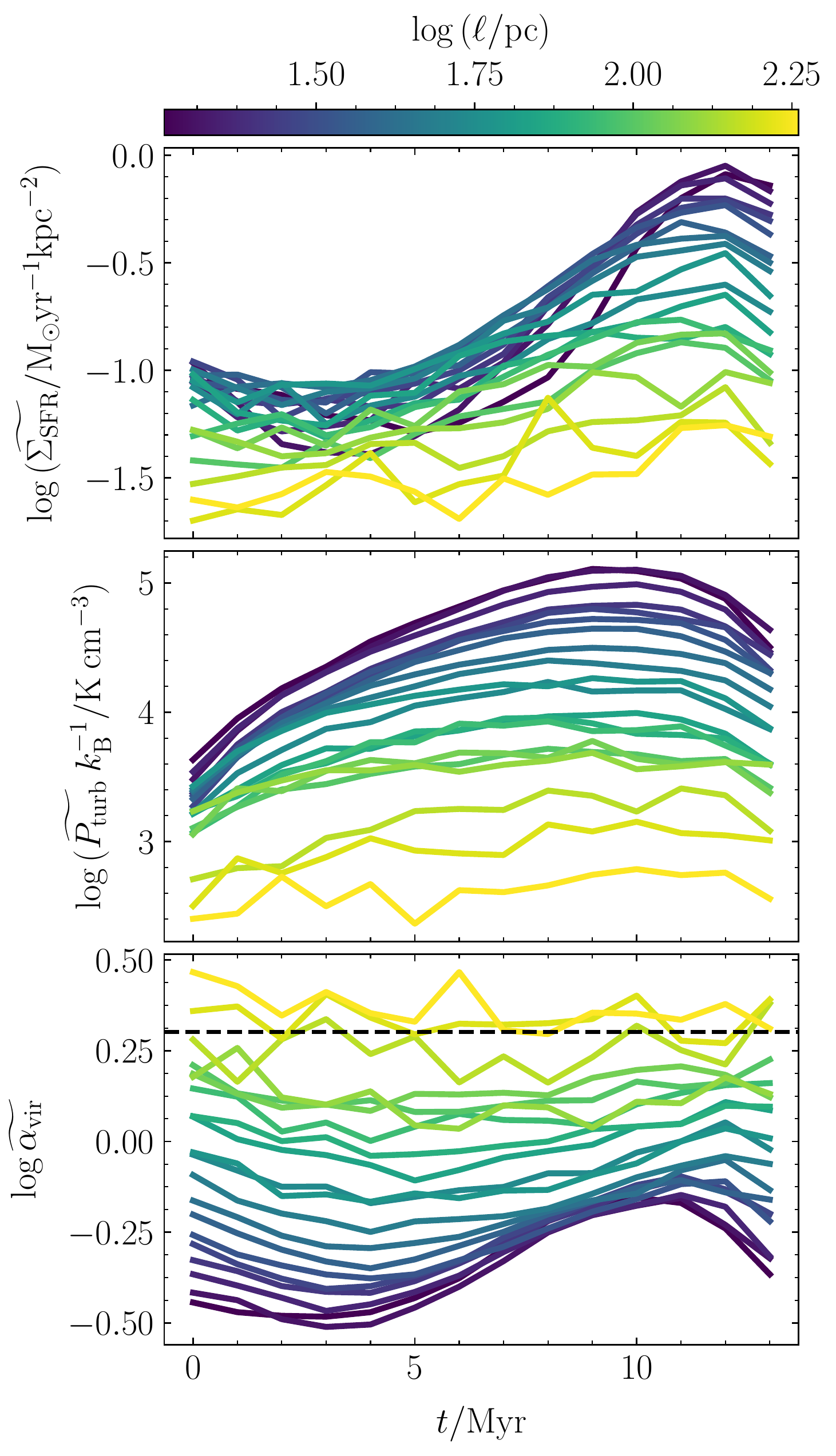}
  \caption{Median values of the star formation rate surface density (upper panel), turbulent pressure (centre panel) and virial parameter (lower panel) as a function of the time $t$ after formation for molecular clouds that survive for $13$~Myr. Values of the star formation rate surface density equal to zero are excluded. The cloud sample is aggregated across all spatial resolutions for the three simulated galaxies and is divided into bins of cloud spatial scale $\ell$, as indicated by the line colour. A virial parameter of $\alpha_{\rm vir} = 2$ (below which the clouds are approximately gravitationally-bound) is given by the black dashed horizontal line.}
\end{figure}

\subsection{The characteristic molecular cloud lifetime, $\tau_{\rm life}$} \label{Sec::destruction-timescale}
Our walk through the cloud evolution network yields a distribution of lifetimes $t_{\rm life}$ for every cloud identified in our simulations, corresponding to the lengths of unique trajectories in Figure~\ref{Fig::merger-tree}. The range of possible lifetimes across all scales spans from $t_{\rm life} = 1$~Myr (the temporal resolution of the network) up to $t_{\rm life} = 120$~Myr, where the longest-surviving clouds undergo many mergers and splits throughout their lifetimes. The distribution $D(t_{\rm life}>t)$ of the number of clouds with lifetimes equal to or longer than time $t$ is shown in the top panel of Figure~\ref{Fig::cloud-lifetimes}, for different cloud scales $\ell$.\footnote{We note that the cloud scale may change as a function of time along a given trajectory. As such, the quoted values of $\ell$ correspond to the median of the time-averaged cloud size, where the median is computed across the cloud population.} Its exponential form is expected for a system in which the characteristic rate of cloud formation $\xi_{\rm form}$, and the characteristic lifetime $\tau_{\rm life}$ before cloud destruction, are time-invariant.\footnote{For our simulations, this assumption is valid: over a period of $400$~Myr, the global galactic SFR changes by just $0.5 \: {\rm M}_\odot \: {\rm yr}^{-1}$ and the size of the cloud population varies by just 1~per~cent.} In this case, the number of clouds $N_{\rm cl}$ in the population obeys the rate equation
\begin{equation} \label{Eqn::rate-equation}
\frac{\dd N_{\rm cl}(t)}{\dd t} = -\tau_{\rm life}^{-1} N_{\rm cl}(t) + \xi_{\rm form}.
\end{equation}
Integrating yields the time-dependence of the population size as
\begin{equation} \label{Eqn::integrated-rate-equation}
N_{\rm cl}(t) = \tau_{\rm life} \xi_{\rm form} + (N_{\rm cl,0} - \xi_{\rm form}\tau_{\rm life}) \exp{\Big[-\frac{t}{\tau_{\rm life}}\Big]},
\end{equation}
where $N_{\rm cl,0}$ is the number of clouds at time $t = 0$. We see that at times $t \gg \tau_{\rm life}$ we reach a steady state given by
\begin{equation} \label{Eqn::rate-equation-eqm-soln}
N_{\rm cl}(t\rightarrow \infty) \rightarrow \tau_{\rm life} \xi_{\rm form}.
\end{equation}
Now, the distribution $D(t_{\rm life}>t)$ of the total number of clouds in the network with lifetimes $t_{\rm life}>t$ is equivalent to the distribution of clouds formed at time $t=0$ that survive up to time $t$ (ignoring clouds formed after $t=0$). This, in turn, decays exponentially as
\begin{equation} \label{Eqn::trajectory-dstbn}
D(t_{\rm life}>t) = N_{\rm cl,0} \exp{\Big[-\frac{t}{\tau_{\rm life}}\Big]},
\end{equation}
explaining the form of the distributions in the upper panel of Figure~\ref{Fig::cloud-lifetimes}. We therefore extract the characteristic cloud lifetime $\tau_{\rm life}$ by fitting a linear function to $\ln{D(t_{\rm life}>t)} \propto -t/\tau_{\rm life}$ and calculating the negative inverse of its slope.

\subsection{Scaling relation of the characteristic molecular cloud lifetime} \label{Sec::cloud-lifetimes}
In the lower panel of Figure~\ref{Fig::cloud-lifetimes}, we show the scaling relation of the characteristic molecular cloud lifetime $\tau_{\rm life}$, which varies across the range $\tau_{\rm life}/{\rm Myr} \in [13, 20]$. It is well-described by the piece-wise function
\begin{equation} \label{Eqn::lifetime-scaling-rln}
  \tau_{\rm life}/{\rm Myr} =
  \begin{cases}
   51 (\ell/{\rm pc})^{-0.30 \pm 0.02} & \text{if $\ell < 100$~pc} \\
   13 & \text{if $\ell \ga 100$~pc}.
  \end{cases}
\end{equation}
The break in the scaling relation occurs at approximately the gas-disc scale-height in our simulations, $\ell \sim h_{\rm g} \sim 100$~pc (vertical black line). Below $h_{\rm g}$, the cloud lifetime increases monotonically as the cloud size decreases. Above $h_{\rm g}$, the cloud lifetime holds constant at approximately the gas-disc crossing time in our simulations, $\tau_{\rm life} \sim t_{\rm cross} \sim 13$~Myr (horizontal black line).

In Figure~\ref{Fig::collapse-profiles}, we demonstrate that the break in the scaling relation can be explained by the fact that molecular clouds of sizes larger than or equal to the gas disc scale-height (yellow-green lines) are not significantly self-gravitating. Throughout their lifetimes, they have low median star formation rate surface densities $\Sigma_{\rm SFR}$ (upper panel) and turbulent pressures $P_{\rm turb}$ (central panel), as well as high median virial parameters $\alpha_{\rm vir}$ (lower panel). We have chosen to show a sample of clouds with lifetimes equal to $13$~Myr for this example, but the result holds equally-well for any survival time. As such, clouds identified on scales $\ell > h_{\rm g} \sim 100~{\rm pc}$ are not destroyed by gravitational collapse and stellar feedback, but are simply destroyed on their turbulent crossing times, which are equivalent to the gas disc crossing time, because all such clouds are vertically-confined by the gas-disc scale-height.

\begin{figure}
 \label{Fig::collapse-dispersal-times}
  \includegraphics[width=\linewidth]{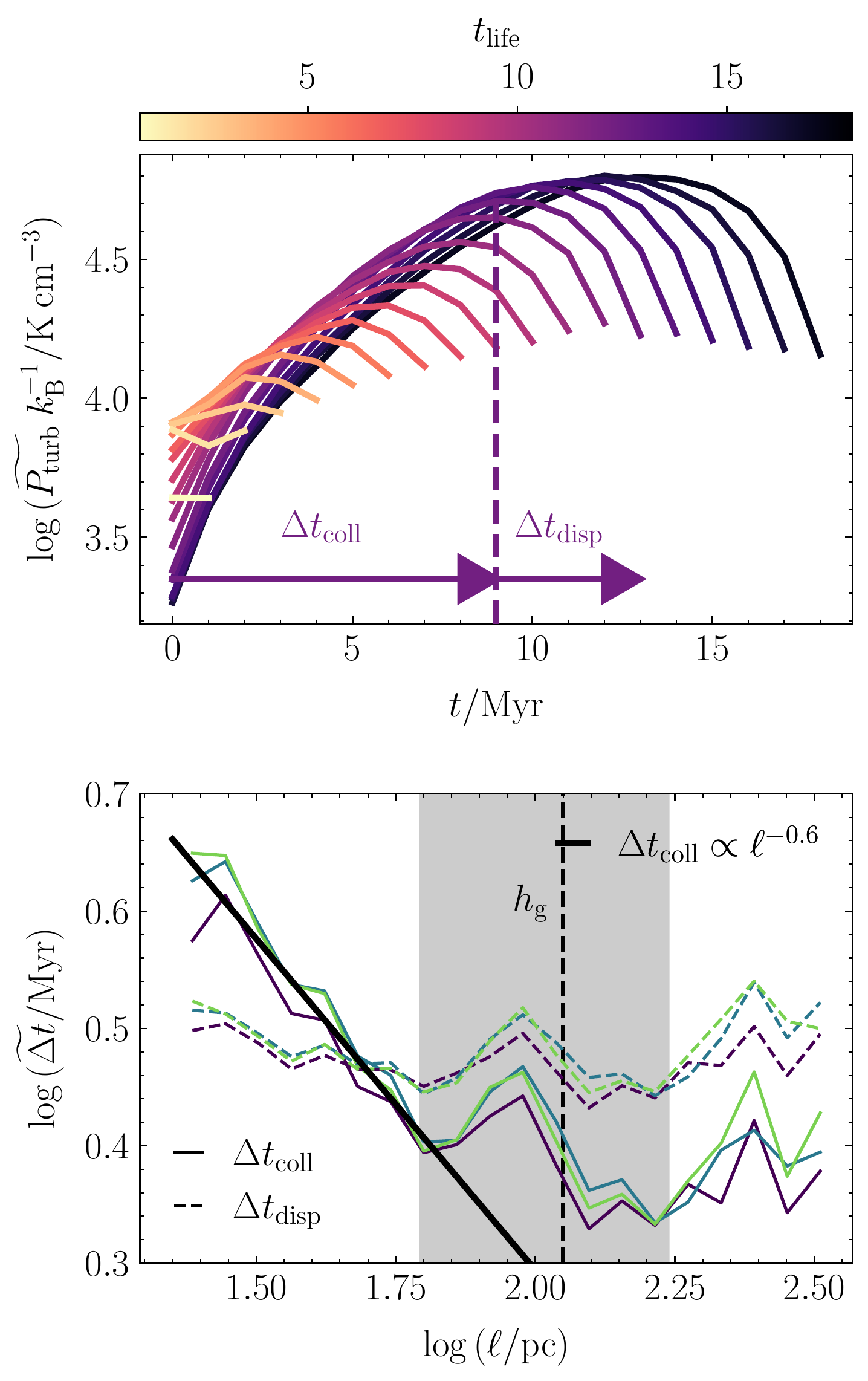}
  \caption{\textit{Top:} Time-evolution of the median turbulent pressure $\widetilde{P_{\rm turb}}$ for clouds surviving for different lengths of time $t_{\rm life}$. The collapse time-scale $\Delta t_{\rm coll}$ (time from cloud formation to maximum pressure) and the feedback dispersal time-scale $\Delta t_{\rm disp}$ (time from maximum pressure to cloud destruction) are indicated for the profile of length $13$~Myr. \textit{Bottom:} Median collapse time-scale $\Delta t_{\rm coll}$ and feedback dispersal time-scale $\Delta t_{\rm disp}$ as a function of the cloud scale $\ell$. The purple, blue and green lines represent the FLAT, SLOPED and CORED simulations, respectively. The black solid line denotes the power-law fit to the combined data for the collapse time-scale across all three simulations. The dashed black line gives the average gas disc scale-height, similarly to Figure~\ref{Fig::cloud-lifetimes}.}
\end{figure}

By contrast, the blue-purple lines in Figure~\ref{Fig::collapse-profiles} demonstrate that at scales $\ell < h_{\rm g} \sim 100~{\rm pc}$, the identified molecular clouds are more likely to be self-gravitating. They collapse to a state of maximum boundedness and turbulent pressure, accompanied by an increase in the star formation rate surface density. After the star formation rate has reached its maximum value, the clouds experience a subsequent decrease in boundedness and pressure that continues until their deaths, consistent with the injection of turbulent kinetic energy by star formation feedback. The collapse times $\Delta t_{\rm coll}$ and dispersal times $\Delta t_{\rm disp}$ are examined explicitly in Figure~\ref{Fig::collapse-dispersal-times}. The upper panel shows the evolution of the median turbulent pressure for clouds surviving for different lengths of time. We see that the longer a cloud survives, the greater the extent of its gravitational collapse to a high turbulent pressure. When stellar feedback sets in, the turbulent pressure drops rapidly as the cloud is unbound and dispersed. In the lower panel, we show that while the dispersal time is approximately-constant across all cloud scales, the collapse time increases as $\Delta t_{\rm coll} \propto \ell^{-0.6}$ for clouds below the gas-disc scale-height. To explain this, we recall the subtle point that the `scale' assigned to each cloud is in fact the average (median) scale over its lifetime. If a cloud is gravitationally-collapsing (as is the case for clouds with $\ell < h_{\rm g} \sim 100~{\rm pc}$), the median size decreases as collapse progresses. Therefore, for an observed population of collapsing clouds, smaller clouds are denser, with shorter instantaneous free-fall times (as shown in Figure~\ref{Fig::collapse-profiles}), but they are more likely to have evolved for a longer period to reach their current state. This means that they are more likely to have longer lifetimes, in accordance with Figure~\ref{Fig::cloud-lifetimes}.

\subsection{Comparison of the cloud lifetime to that measured with the method of~\protect\cite{Kruijssen18a}} \label{Sec::sim-vs-heisenberg}
In the preceding sub-section, we showed that the cloud lifetime obeys a power-law scaling relation below the gas-disc scale-height $h_{\rm g}$, and argued that this trend is driven by gravitational collapse and the subsequent dispersal of clouds by stellar feedback. In this section, we focus on the lifetimes of clouds identified at scales larger than or equal to $h_{\rm g}$. These objects are approximately gravitationally-unbound and vertically-confined by the disc scale-height, and so are dispersed on the gas disc crossing time $t_{\rm cross}$. In this section, we show that the position of the break in the scaling-relation ($\sim h_{\rm g}$) and the lifetimes of clouds above this break ($\sim t_{\rm cross}$) can alternatively be obtained by applying a statistical model for the gas-to-stellar flux ratio on different scales~\citep{Kruijssen2014,Kruijssen18a} to the simulated molecular gas and SFR column densities from our simulations. This method has so far been applied to a range of direct extragalactic observations~\citep{Kruijssen2019,Chevance20,2020arXiv201200019K,Ward20,Zabel20}.

\begin{figure}
 \label{Fig::sim-vs-heisenberg}
  \includegraphics[width=\linewidth]{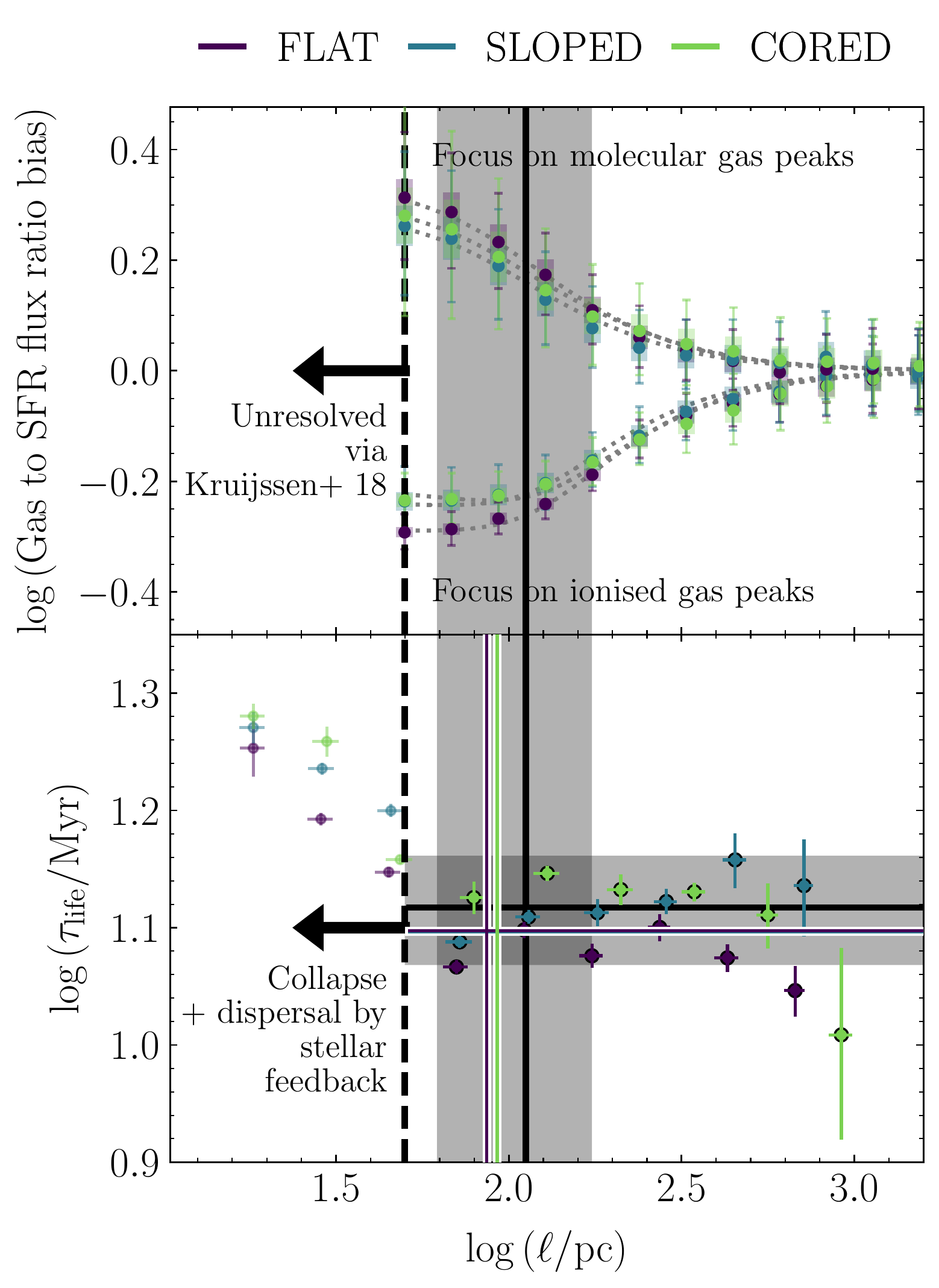}
  \caption{\textit{Top panel:} The gas-to-SFR flux ratio relative to the galactic average value as a function of map resolution (minimum resolvable cloud size) for each simulated galaxy at $t=600$~Myr. The upper branch represents apertures focussed on molecular gas peaks, while the lower branch represents apertures focussed on stellar surface density peaks for stars with ages $<5$~Myr. The dotted lines show the best-fitting models using the method of~\protect\cite{Kruijssen18a}, and the thick dashed vertical line represents the maximum-resolution map used in the application of this method. \textit{Bottom panel:} The scaling relation for the molecular cloud lifetime as presented in Figure~\protect\ref{Fig::cloud-lifetimes}, but including the time- and disc-averaged values of the gas-disc scale-height (vertical black line) and crossing time (horizontal black line) across our simulations. The separation length and gas phase duration derived using the method of~\protect\cite{Kruijssen18a} are given by the white-bordered vertical and horizontal lines, respectively.}
\end{figure}

The model of~\cite{Kruijssen18a} fits the bias of the gas-to-young stellar\footnote{The stellar surface density is computed for an age bin of $<5$~Myr.} flux ratio away from the galactic average value, within apertures of variable size $l_{\rm ap}$ centred on peaks of gas emission (upper arm in the top panel of Figure~\ref{Fig::sim-vs-heisenberg}) or on peaks of young stellar emission (lower arm). The model is parametrised by $l_{\rm ap}$, a mass-weighted mean separation length $\lambda_{\rm K18}$ of `independent star-forming regions', and a set of mass-weighted mean time-scales spent by these regions in the gas-dominated, stellar-dominated and combined gas-stellar phases of star formation. The regions are therefore ‘independent’ in the sense that they evolve independently through the star-forming phases, so that the evolutionary stages of neighbouring regions are uncorrelated.

For our simulations, the model of~\cite{Kruijssen18a} is fitted to maps of the gas and stellar column densities, spaced at $50$~Myr intervals between simulation times of $600$~Myr and $1$~Gyr. The value of $\lambda_{\rm K18}$ is averaged over time for each simulation, and is given by the white-bordered vertical lines in the lower panel of Figure~\ref{Fig::sim-vs-heisenberg}. Similarly, the total duration $\tau_{\rm K18}$ of the gas phase is obtained by summing the time-scales of the gas-dominated and combined gas-stellar phases, then taking the time-average of the result, indicated by the white-bordered horizontal lines in the lower panel of Figure~\ref{Fig::sim-vs-heisenberg}. The maps have pixels of size $30$~pc and are convolved to resolutions across the range $50~{\rm pc} < l_{\rm ap} < 4~{\rm kpc}$. The highest map resolution is indicated by the thick dashed vertical line in Figure~\ref{Fig::sim-vs-heisenberg}.

In the lower panel of Figure~\ref{Fig::sim-vs-heisenberg}, we show that the separation length $\lambda_{\rm K18}$ derived via the method of~\cite{Kruijssen18a} is consistent with the position of the break in the scaling relation of the cloud lifetime, which in turn is consistent with the average gas-disc scale-height $h_{\rm g} \sim 112 \pm 50$~pc (solid black vertical line and grey-shaded region). Similarly, the gas-phase duration $\tau_{\rm K18}$ is consistent with the value of the cloud lifetime above the break, which in turn is consistent with the average gas-disc crossing time $t_{\rm cross} \sim 13.1 \pm 0.6$~Myr (solid black horizontal line and grey-shaded region).

This result can be understood as follows. Gravitationally-unbound regions of vertical extent $h_{\rm g}$ are shaken apart by turbulence on the gas disc crossing time $\tau_{\rm K18} \sim \tau_{\rm life}(\ell \ga  h_{\rm g}) \sim t_{\rm cross}$. Within the galactic mid-plane, communication between such regions therefore breaks down at a scale $h_{\rm g}$. That is, `independent regions' in the sense of~\cite{Kruijssen18a} are separated by a length-scale of $\lambda_{\rm K18} \sim h_{\rm g}$. This is therefore also the length-scale below which the gas and stellar fluxes de-correlate from the galactic average value, as shown in the top panel of Figure~\ref{Fig::sim-vs-heisenberg}. Objects identified at scales $\ell \ga h_{\rm g}$ can be interpreted as unresolved collections of such `independent regions'.

We note that this result is consistent with panel (c) of Figure 2 in~\cite{Kruijssen2019}, which shows close agreement between the separation length $\lambda_{\rm K18}$ in NGC 300 and the gas-disc scale-height. Similarly, Figure 5 of~\cite{Chevance20} shows close agreement between the gas phase duration $\tau_{\rm K18}$ and the gas-disc crossing time (equivalently the cloud crossing time on cloud scales $h_{\rm g}$) for eight out of nine nearby galaxies.

\begin{figure*}
 \label{Fig::betaQ-lifetime}
  \includegraphics[width=\linewidth]{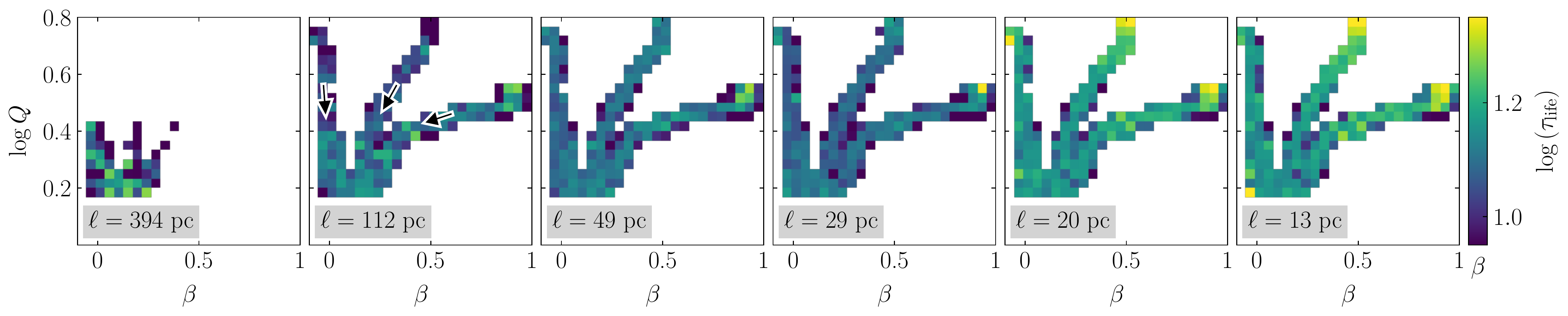}
  \caption{Characteristic cloud lifetime $\tau_{\rm life}$ for populations of clouds across different galactic-dynamical environments. The value of $\tau_{\rm life}$ (coloured pixel in each panel) is obtained via the exponential distribution of trajectory lengths in each bin of the shear parameter $\beta = \dd \ln{v_c(R)}/\dd R$ for a galactic circular velocity of $v_c(R)$ (horizontal axis) and the Toomre $Q$ stability parameter (vertical axis), via the method of Section~\ref{Sec::destruction-timescale}. The data from the cloud evolution networks of all three simulated discs is compiled for each spatial resolution $\epsilon = 198$~pc through to the native resolution of $6$~pc (top left panel to lower right panel). The median cloud scale $\ell$ for each cloud evolution network is annotated in the grey-shaded boxes, and each distinct set of connected pixels corresponds to the locus of the cloud population for one isolated disc galaxy. The black arrows mark the direction in which the galactocentric radius increases.}
\end{figure*}

\subsection{Variation of the cloud lifetime with the galactic environment} \label{Sec::betaQ_lifetimes}
Given the universally self-gravitating behaviour of our simulated molecular clouds below the scale-height of the galactic disc, we do not expect that the characteristic molecular cloud lifetime $\tau_{\rm life}$ will depend on the galactic-dynamical environment at any scale $\ell \la h_{\rm g} \sim 100$~pc. That is, the cloud lifetime varies with the time-scale $\Delta t_{\rm coll}$ for gravitational contraction and with the time-scale $\Delta t_{\rm disp}$ for dispersal by means of star formation, which are local quantities that depend on the cloud density and on the physics of stellar feedback, and not on the larger-scale properties of the galaxy. In Figure~\ref{Fig::betaQ-lifetime} we verify our suspicion by examining the characteristic cloud lifetime as a function of the galactic-dynamical environment across our three simulated galaxies. Each of the six panels corresponds to the combined population of molecular clouds in the FLAT, SLOPED and CORED simulations, identified at spatial resolutions from $\epsilon = 198$~pc (top left panel) to the native resolution of $\epsilon = 6$~pc (bottom right panel). The grey boxes display the median cloud scale $\ell$ within each map. Across all resolutions and scales, no clear colour gradient is visible, indicating that there is no appreciable trend with the galactic-dynamical environment. This result is in agreement with the finding of~\cite{Jeffreson20} that the molecular clouds in these Milky Way-pressured simulations are highly over-dense and over-pressured relative to the galactic mid-plane, such that their turbulent and star-forming properties are decoupled from galactic dynamics, and driven instead by local gravitational effects. As noted in~\cite{Jeffreson20}, galactic dynamics might become important in galaxies with higher mid-plane gas pressures, but confirmation of this suspicion would require further explicit investigation. We note that we would expect some small degree of variation of $\tau_{\rm life}$ with the galactocentric radius (indicated by the black arrows in the lower left-hand panel) at a scale of $198$~pc, due to the variation of the gas-disc scale-height and thus the gas-disc crossing time. Unfortunately there is insufficient data across environments to distinguish such a trend at this low resolution.

\section{Discussion} \label{Sec::discussion}
\subsection{Comparison to simulations from the literature}
The identification of distinct molecular clouds in our simulations allows for a comparison to the distributions of cloud lifetimes derived in similar numerical simulations from the literature~\citep{DobbsPringle13,Fujimoto19,Dobbs19,Benincasa2019}. Two major differences in our approach relative to these works, which may significantly influence the comparison of our derived cloud lifetimes, are discussed below. These are (1) our choice of star formation efficiency, and (2) our cloud-tracking procedure.

A major unknown influence on our derived cloud lifetimes is our choice of star formation efficiency: $\epsilon_{\rm ff} = 1$~per~cent above our star formation threshold, $n_{\rm thresh} = 1000~{\rm cm}^{-3}$. As discussed in Section~\ref{Sec::simulations}, our value of $\epsilon_{\rm ff}$ is motivated by observations of the average star formation efficiency across the interstellar medium. However, within the densest parts of molecular clouds (and therefore at volume densities $n_{\rm H} > n_{\rm thresh}$), a value closer to 10~per~cent may be more appropriate~\citep{2009yCat..21810321E}. A full investigation of the $\epsilon_{\rm ff}$-variation in our results is beyond the scope of the current work, however we might naively assume that a higher value would lead to shorter molecular cloud lifetimes, and to a reduction in their environmental-dependence~\citep{Dobbs11,Semenov18,Semenov19}. A range of values of $\epsilon_{\rm ff}$ are used in the literature: $\epsilon_{\rm ff}=1$~per~cent in~\cite{Fujimoto19}, $20$~per~cent in~\cite{Dobbs19} and $100$~per~cent for locally self-gravitating gas in~\cite{Benincasa2019}. However, it is impossible to draw concrete conclusions from a comparison of the cloud lifetimes obtained in these works, due to substantial differences in the stellar feedback models used.

Unlike the cloud-tracking procedures used in~\cite{DobbsPringle13,Fujimoto19,Dobbs19,Benincasa2019}, we pick out molecular clouds in two spatial dimensions (rather than three) and follow these clouds via the Eulerian (rather than Lagrangian) flow of gas mass. As discussed in Section~\ref{Sec::cloud-ID}, our procedure is more closely-comparable to direct observations, which commonly identify clouds in position-position-velocity space~\citep[e.g.][]{Sun18,Sun2020}. In addition, we have explicitly checked the three-dimensional structure of the clouds in our sample by examining the distribution of the CO-luminous gas (used to compute all physical cloud properties in this work) along the line-of-sight ($z$-axis). We find that our sample contains $<2$~per~cent of clouds across all resolutions ($<6$~per~cent at the highest resolution $\epsilon = 6$~pc) with more than 10~per~cent of their CO-luminous gas mass in structures that overlap in the galactic plane, but are separated by more than $\epsilon$ along the $z$-axis (line-of-sight). This indicates that, as well as being closely-comparable to observational cloud-identification techniques, our method is reliable to better than 95~per~cent (in cloud mass) at picking out three-dimensional clouds using just the two-dimensional distribution of CO-luminous gas. In the future, it will be interesting to include the gas velocity data in our cloud-identification procedure, to more-closely match direct position-position-velocity observations of molecular gas. In comparison to other simulations in the literature,  we find that our range of characteristic lifetimes $13~{\rm Myr} \la \tau_{\rm life} \la 20~{\rm Myr}$ across the scale range $10~{\rm pc} \la \ell \la 1~{\rm kpc}$ is comparable to the typical span of $4$-$25$~Myr found by~\cite{DobbsPringle13} at a similar numerical resolution. At our largest spatial scales $\ell \ga 100$~pc, our values are still around twice the mean cloud lifetimes measured by~\cite{Benincasa2019}, however their mass resolution is around ten times lower than ours, and so this effect may be attributed to missing the lower-mass `tails' of cloud formation and destruction that we resolve. Our range of lifetimes is significantly shorter than the mean values of $30$-$40$~Myr computed by~\cite{Fujimoto19}, however as studied in detail by these authors, their elongated cloud lifetimes are likely due to inefficient stellar feedback in their simulated discs.

Our discussion of the merger rate for giant molecular clouds is closely-related to work by~\cite{Dobbs15}, who have also constructed cloud evolutionary networks in order to characterise cloud interactions as a function of time. In their flocculent disc galaxy simulation, they find a merger rate of one in $\sim 28$~Myr for clouds with diameters of $\ell \sim 100$~pc; comparable to the values we obtain at similar scales. At smaller $\ell$, we obtain cloud merger rates almost three times faster, up to one in every $10$-$12$~Myr. This regime is not examined by~\cite{Dobbs15}, who take a stricter threshold for cloud identification, allowing only those structures containing $50$ or more gas cells, with masses $M>1.5 \times 10^4~{\rm M}_\odot$. By contrast, we have allowed clouds all the way down to a few solar masses, containing only five to ten Voronoi cells in some cases. Our lenient cloud identification threshold is chosen to ensure the consistency of our cloud-tracking procedure across maps at different spatial resolutions, and it is validated to an extent by the lack of a spurious small-scale turnover in the scaling relations we derive. However, a result of our leniency may be an elevated frequency of `mergers' occurring between clouds of very different masses (i.e.~very low-mass clouds with very high-mass clouds). These events might better be considered as accretion events and excluded from the merger sample. In this work we have remained as agnostic as possible towards definitions of mergers and splits via their mass ratios, but in future work this could easily be incorporated into the MC random walk described in Section~\ref{Sec::random-walk} by means of a non-uniform MC sampling criterion.

A particular point of agreement between our work and that of~\cite{Dobbs15} is that cloud interactions, though frequent, have little appreciable effect on the internal turbulent or star-forming properties of the interacting clouds. Although it is tempting to conclude that cloud interactions have no effect on the galactic star formation rate, we must be careful to state the caveat that neither our simulations (at mass resolution $\sim 900~{\rm M}_\odot$), nor those of~\citealt{Dobbs15} (at mass resolution $\sim 300~{\rm M}_\odot$), explicitly resolve star formation, instead relying on a parametrisation of the empirical star formation relation~\citep[][see our Equation~\ref{Eqn::starformation}]{Kennicutt98}. This means that the star formation resulting from slow gravitational collapse will be well-characterised in our simulations, but gas that is bumped into the high-density regime at shorter time-scales, as in shocks, may not be properly modelled. Simulations of discrete colliding clouds at high spatial resolution do indeed find an elevation of the star formation efficiency owing to the formation of filamentary structures and sheets on sub-cloud scales~\citep{2014ApJ...792...63T,2015MNRAS.453.2471B,Balfour17,2017ApJ...841...88W,2020MNRAS.494..246T}. Similarly, previous simulations have investigated colliding flows driven by magnetohydrodynamic turbulence~\citep[e.g.][]{1995ApJ...455..536P,1995MNRAS.277..377P,1999ApJ...515..286B,1999ApJ...527..285B,2000A&A...359.1124H,2002ApJ...578..256L,2005MNRAS.359..809C,2005ApJ...633L.113H,2006ApJ...648.1052H,2014ApJ...793...84Z} or due to expanding bubbles driven by stellar feedback~\citep[e.g.][]{1995ApJ...440..634R,1999ApJ...514L..99K,Slyz2005,2005ApJ...626..864M,KimCG&Ostriker15a,KimCG&Ostriker15b}, which in theory should operate and trigger star formation on all levels of the interstellar medium hierarchy examined in this work~\citep{1973PASJ...25....1S,1991ApJ...378..139E,Elmegreen93,1996ApJ...471..816E,2007ApJ...668.1064E}. Such simulations find continuous velocity fields that cut across the boundaries of discrete, identified clouds, indicating the presence of converging flows at their edges. At our resolutions, no such triggered star formation is observed, but we cannot rule out its presence at higher resolutions. Ultimately, both a large statistical sample of clouds like the one presented here, plus sufficient numerical resolution to resolve shocks at the interfaces of converging flows and cloud interactions, is required to rule out such effects. This could possibly be achieved using zoom-in simulations of cloud samples from a larger isolated galaxy simulation.

Finally, we have found in this work that self-gravitating clouds (those below the gas-disc scale-height) collapse to a maximum density of star formation, and then are dispersed (likely by stellar feedback from massive stars). This finding is consistent with the work of~\cite{Semenov17,Semenov18} who show that the long depletion times in galaxies are due to the cycling of gas between the dense, cold and supersonic star-forming phase (corresponding to the clouds smaller than the gas-disc scale-height in our simulations) and the diffuse, warm, sub-sonic phase (dominating the masses and volumes of the clouds we identify above the gas-disc scale-height). These authors follow parcels of gas through cycles of collapse, star formation and dispersal, demonstrating that only a small fraction of molecular gas is converted to stars during each cycle. We have therefore shown that in order to characterise the time-scales on which these cycles of collapse and dispersal occur at the highest-density levels of the hierarchical interstellar medium, observations must resolve scales significantly below the scale-height of the galactic gas disc.

\subsection{Comparison to observations from the literature}
Our findings are compatible with the age-spreads of Cepheid variables and stellar clusters observed by~\cite{1996ApJ...466..802E,1998MNRAS.299..588E,2000ApJ...530..277E}. These observations demonstrate that star formation occurs on $1$-$2$ crossing times across two orders of magnitude in spatial scale, from $10$~pc up to $\sim 1$~kpc. Although our cloud lifetimes decrease with increasing spatial scale below the gas disc scale-height, while the crossing time increases, the density of star formation peaks at the time of maximum collapse (and therefore at the smallest cloud size). This means that at the smallest scales, regions of star formation are most likely to be temporally-separated by the instantaneous free-fall time, and not by the preceding period of cloud evolution. Therefore, in both regimes (larger than and smaller than the gas disc scale-height), we find that star formation occurs within approximately $1$-$2$ cloud crossing times (driven by the turbulent crossing time, and by the free-fall time, respectively).

We may also compare our numerically-derived cloud lifetimes to observed values from nearby galaxies. The lack of temporal information in direct observations means that these values have been determined either by (1) measuring the velocities and separations of clouds that are assumed to form part of an evolutionary sequence~\citep{ScovilleHersh1979,Solomon1979,Engargiola03,Meidt15}, or (2) using the numbers of clouds in different evolutionary phases as a proxy for the time intervals spent in these phases~\citep{Blitz2007,Kawamura09,Murray11,Corbelli17}. In Milky Way-mass galaxies, these studies generally yield cloud lifetimes in the range $10$-$30$~Myr, in agreement with our simulated values. In addition, we have discussed in Section~\ref{Sec::sim-vs-heisenberg} that the cloud lifetimes we obtain at and above the gas disc scale-height are consistent with the gas disc crossing time for our simulated galaxies, in agreement with the cloud lifetimes derived on similar scales from observations of nearby galaxies~\citep{Kruijssen2019,Chevance20}.

\section{Conclusions} \label{Sec::conclusion}
In this work, we have examined the time-evolution of giant molecular clouds across Milky Way-like environments, using a set of three isolated galaxy simulations in the moving-mesh code {\sc Arepo}. The galaxies are designed to probe a wide range of galactic-dynamical environments, spanning an order of magnitude in the Toomre $Q$ gravitational stability parameter, the galactic orbital angular velocity $\Omega$, and the mid-plane hydrostatic pressure~\citep{Jeffreson20}, as well as the full range of galactic shear parameters $\beta$ from the case of solid-body rotation ($\beta = 1$) up to the case of a flat rotation curve ($\beta = 0$). We have found that:
\begin{enumerate}
 \item The cloud evolutionary network of each galaxy is highly-substructured in space and in time. Around $80$~per~cent of clouds at spatial scales of $\ell = 10$-$20$~pc interact with other clouds during their lifetimes, with a merger rate of $\Gamma_{\rm merge} \sim 0.1~{\rm Myr}^{-1}$. The rate drops to one in thirty at cloud scales of $\ell \sim 400$~pc.
 \item The merger rate is well-described by the crossing time in a supersonically-turbulent, fractally-structured interstellar medium, with a fractal index of $D \sim 1.54$. This relationship depends on the two-dimensional velocity dispersion $\sigma_{\lambda, {\rm 2D}}$ of molecular cloud centroids within the galactic mid-plane, which is found to obey the same scaling relation with cloud separation $\lambda$ as is obeyed by the three-dimensional internal cloud velocity dispersion $\sigma_{\rm 3D}$ with cloud scale $\ell$~\citep{Larson1981,Heyer+09}. This correspondence extends up to scales ten times larger than the gas disc scale-height. That is, supersonic turbulence sets the two-dimensional structure in the molecular gas of our galaxies over a scale range of $10~{\rm pc} \la \lambda \la 1$~kpc, in agreement with~\cite{2000ApJ...530..277E,2003ApJ...590..271E,2003ApJ...593..333E}.
 \item Despite the frequency of cloud mergers, they do not appear to significantly alter the physical properties of the molecular clouds in our simulations. As clouds pass through mergers, their star-forming and turbulent properties continue to evolve as they did before the merger.
 \item However, clouds that undergo mergers or splits during their lifetimes display small systematic differences in their physical properties, relative to those that evolve in complete isolation. A study of the demographics of these two cloud populations is a topic for future work.
 \item The distribution of molecular cloud lifetimes in each galaxy takes an exponential form with values between $1$ and $120$~Myr, indicating that the cloud population $N_{\rm cl}$ is well-described by a rate equation of the form
\begin{equation}
\frac{\dd N_{\rm cl}}{\dd t} = \tau_{\rm life}^{-1} N_{\rm cl} + \xi_{\rm form},
\end{equation}
 where $\xi_{\rm form}$ is the rate of cloud formation and $\tau_{\rm life}$ is the characteristic cloud lifetime for the population (the characteristic time-scale of cloud destruction).
 \item We find that $\tau_{\rm life}$ obeys a scaling relation of the form $\tau_{\rm life} \propto \ell^{-0.3}$ across all three galaxies below the gas disc scale-height, driven by the competition between gravitational contraction and stellar feedback. Above the scale-height, the characteristic lifetime is constant and set by the crossing time of the galactic disc ($\sim 13$~Myr), in agreement with observations~\citep{Kruijssen2019,Chevance20}. The range of characteristic lifetimes across spatial scales is $13~{\rm Myr} \la \tau_{\rm life} \la 20~{\rm Myr}$.
 \item Below the gas-disc scale-height, the simulated populations of molecular clouds are self-gravitating and their lifetimes are consequently independent of the galactic-dynamical environment.
\end{enumerate}

\section*{Acknowledgements}
We thank the anonymous referee for an attentive report that improved the presentation of the results in our manuscript. We thank Volker Springel for providing us access to Arepo. SMRJ is supported by Harvard University through the ITC. We gratefully acknowledge funding from the Deutsche Forschungsgemeinschaft (DFG, German Research Foundation) through an Emmy Noether Research Group (SMRJ, MC, JMDK; grant number KR4801/1-1) and the DFG Sachbeihilfe (MC, JMDK; grant number KR4801/2-1), as well as from the European Research Council (ERC) under the European Union's Horizon 2020 research and innovation programme via the ERC Starting Grant MUSTANG (SMRJ, BWK, JMDK; grant agreement number 714907). SMRJ, MC, JDMK, MRK and YF acknowledge support from a UA-DAAD grant. BWK and AJW acknowledge funding in the form of Postdoctoral Research Fellowships from the Alexander von Humboldt Stiftung. MRK acknowledges support from the Australian Research Council through Future Fellowship FT80100375 and Discovery Projects award DP190101258. The work was undertaken with the assistance of resources and services from the National Computational Infrastructure (NCI; award jh2), which is supported by the Australian Government. 

\section*{Data Availability Statement}
The data underlying this article are available in the article and in its online supplementary material.

\bibliographystyle{mnras}
\bibliography{bibliography}

\bsp

\appendix

\begin{algorithm}
\caption{A single Monte Carlo iteration of the algorithm used to extract the cloud lifetime $\tau_{\rm life}$ from the cloud evolution network. A worded description of the algorithm is given in the text.} \label{alg::cloud-lifetime}
\begin{algorithmic}[1]
\State{$\mathcal{F} = \{f_i\} \gets \text{set of unique formation nodes}$.}
\State{$\Delta t \gets \text{time interval between consecutive nodes}$.}
\vspace*{.3cm}

\State{$\mathcal{R} = \{r_i\} \in U(0,1) \gets \text{set of random numbers for all nodes}$.}
\State{$\mathcal{I} = \{I_i = 0\} \gets \text{no.~of times that each node has been accessed}$.}
\vspace*{.01cm}

\For {$f \text{ in } \mathcal{F}$}
\State $N_f = \theta_{\rm child}(f) - \theta_{\rm par}(f)$
\For {$j = 0 \to N_f$}
\State{$\tau_{\rm life} = {\Call{NextStep}{f,0}}$}
\EndFor
\EndFor
\vspace*{.3cm}

\Function{NextStep}{$n, \tau_{\rm life}$}
\State{$\mathcal{C}_n \gets \text{children of node }n; \; \theta_{\rm child}(n) \equiv |\mathcal{C}_n|$.}
\State{$\mathcal{P}_n \gets \text{parents of node }n; \; \theta_{\rm par}(n) \equiv |\mathcal{P}_n|$.}
\vspace*{.3cm}

\State{$N_{\rm outcomes} \gets \text{total no.~of MC outcomes at }n$.}
\State{$N_{\rm outcomes} = \max{[\theta_{\rm par}(n), \theta_{\rm child}(n)]}$.}
\vspace*{.3cm}

\State{$N_{\rm term} \gets \text{no.~outcomes that result in path termination at }n$.}
\State{$N_{\rm term} = \max{[0, \theta_{\rm par}(n) - \theta_{\rm child}(n)]}$.}
\vspace*{.3cm}

\State{$k = 0$.}
\While{$r_n > k/N_{\rm outcomes}$}
\State{$k = k + 1$.}
\EndWhile
\State{$k = (k + I_n) \mod N_{\rm outcomes}$.}
\State{$I_n = I_n + 1$.}
\If{$k < N_{\rm term}$}
\State{\Return{$\tau_{\rm life}$}}
\Else
\State{$\tau_{\rm life} = \tau_{\rm life} + \Delta t$.}
\State{\Return{\Call{NextStep}{$\mathcal{C}_n[k - N_{\rm term}], \tau_{\rm life}$}}}
\EndIf
\EndFunction
\end{algorithmic}
\end{algorithm}

\section{Calculation of cloud lifetimes} \label{App::cloud-lifetimes}
\begin{figure}
 \label{Fig::schematic-FD}
  \includegraphics[width=\linewidth]{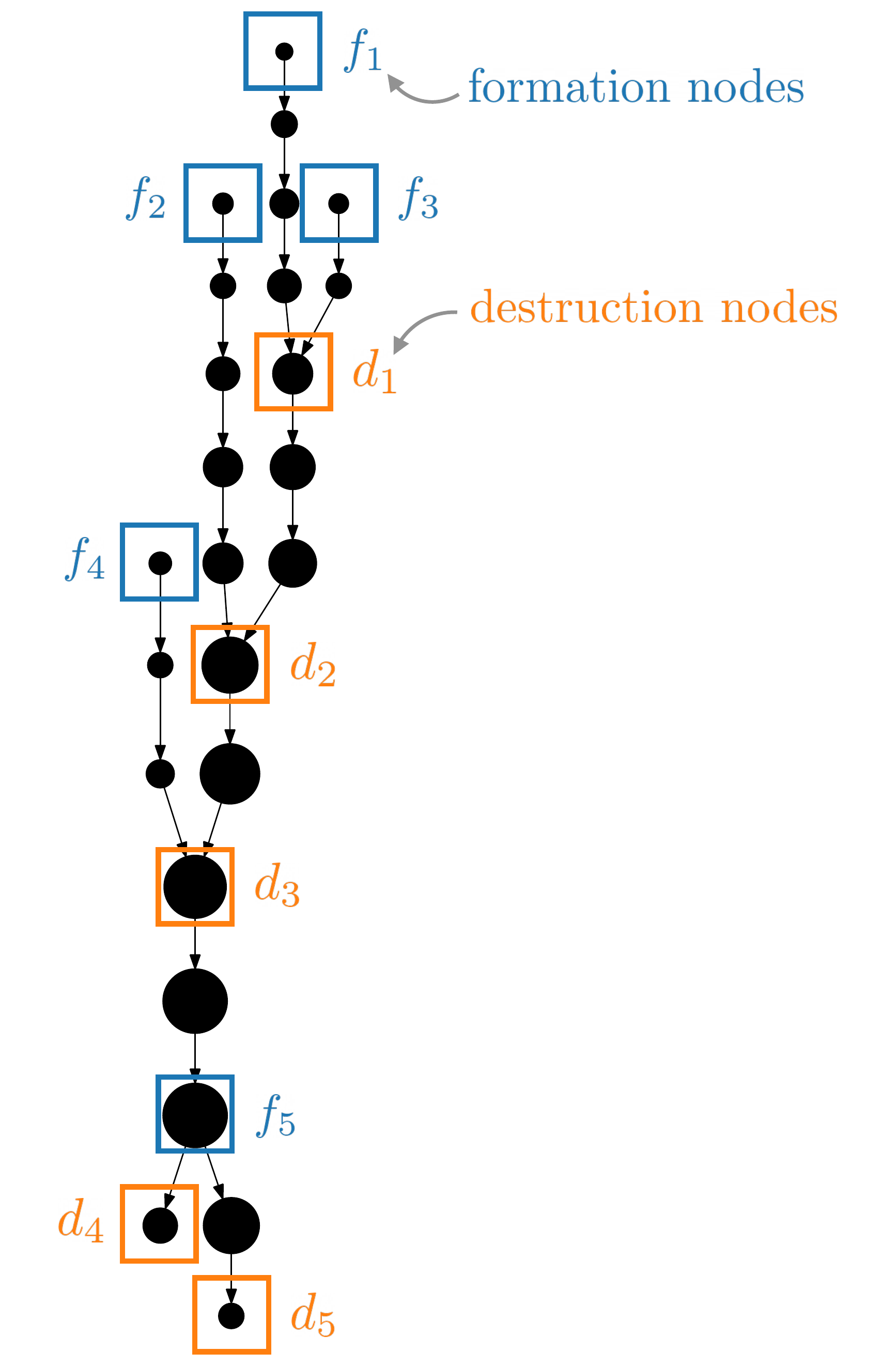}
  \caption{Schematic illustrating the positions of the cloud-formation nodes $f_i; \: i = 1...5$ and cloud-destruction nodes $d_j; \: j = 1...5$ in a single component of the FLAT cloud evolution network. Formation nodes generate a net increase in cloud number, while destruction nodes correspond to a net decrease.}
\end{figure}

As described in Section~\ref{Sec::cloud-lifetime}, we extract the cloud lifetime $\tau_{\rm life}$ from the cloud evolution network according to a Monte Carlo (MC) algorithm that samples from the set of all unique time-directed trajectories through the network. The pseudocode for a single MC iteration is presented in Algorithm~\ref{alg::cloud-lifetime}. During an iteration, a trajectory is sourced at the site of every cloud formation node in the network (see Figure~\ref{Fig::schematic-FD}), and subsequently iterates the cloud lifetime by increments of $\Delta t = 1$~Myr as it steps along edges from parent nodes to their children. At nodes with multiple parents or children ($\theta_{\rm par}>1$ or $\theta_{\rm child}>1$, respectively), we apply an unbiased MC assignment for the path taken by choosing a random number from the uniform distribution $U(0,1)$. Figure~\ref{Fig::schematic-nodetypes} illustrates the MC assignment for different types of nodes, where formation nodes are coloured blue and destruction nodes are coloured orange. The arrow labels indicate the probability of each outcome. At cloud formation nodes $(e)$-$(h)$, all possible MC outcomes correspond to paths for continued cloud evolution, while at cloud destruction nodes $(a)$-$(d)$, a fraction of the possible MC outcomes $(\theta_{\rm par}-\theta_{\rm child})/\theta_{\rm par}$ corresponds to termination of the trajectory. It is also possible to have multiple MC outcomes at nodes such as $(j)$, for which there is no net change in the number of clouds, $\theta_{\rm child}=\theta_{\rm par}$\footnote{At nodes with $\theta_{\rm child}=\theta_{\rm par}$, we could consider a different physical interpretation: that one or all of the incoming clouds are destroyed and re-formed. We choose the interpretation that all clouds survive because the interaction is shorter-lived than the temporal resolution of our simulations, with a duration of $<3$~Myr. At the resolution of our simulations, we therefore have no evidence that a merger has occurred; only that the clouds have interacted and may have exchanged mass.}. By performing 70 MC iterations, we obtain a converged distribution of cloud lifetimes that accounts for all interactions in the cloud evolution network. The procedure satisfies the requirements for cloud number conservation and cloud uniqueness, which are defined for the network as follows:
\begin{enumerate}
 \item \textit{Cloud uniqueness:} Each edge connecting two nodes in the network represents a time-step in the evolution of a single cloud, and so can contribute to just one cloud lifetime. Edges must not be double-counted when calculating cloud lifetimes.
 \item \textit{Cloud number conservation:} Each cloud can be formed and destroyed only once, so the number of cloud lifetimes retrieved from the entire network must be equal to the number of cloud formation events and cloud destruction events.
\end{enumerate}
In the following, we give a detailed worded description of Algorithm~\ref{alg::cloud-lifetime}.
\begin{itemize}
 \item Lines 1-2: Define variables for the entire cloud evolution network, for all MC iterations. A cloud formation node $f_i \in \mathcal{F} = \{f_i\}$ is any node that generates a net increase in the number of clouds, $\theta_{\rm child} > \theta_{\rm par}$. The time interval $\Delta t = 1$~Myr is the time between simulation snapshots, and so between consecutive nodes joined by edges in the network.
 \item Lines 3-4: Define the variables for a single MC iteration. At the beginning of each iteration, we generate a set random number $r_n \in \mathcal{R} = \{r_i\}$ for every node $n$ in the network. At nodes with multiple MC outcomes ($\theta_{\rm child} > 1$ or $\theta_{\rm par} > 1$), this number is used to choose between outcomes. We also keep track of the number of times $I_n \in \mathcal{I} = \{I_i\}$ that node $n$ has been accessed, so that each outcome is accessed exactly once. In this sense, the random number $r_n$ sets the first outcome to be accessed.
 \item Lines 5-8: Loop over the unique cloud formation nodes $f_i \in \mathcal{F}$. Each formation node $f$ sources $N_f$ separate paths, where $N_f$ is the net increase in cloud number generated at $f$. For each separate path initiation, the cloud lifetime is calculated via the recursive function {\sc NextStep} (lines 7-8). In the first call to the function {\sc NextStep}$(f,0)$, the cloud lifetime is initialised to zero.
 \item Line 9: Define the function {\sc NextStep}, taking a node $n$ and a cloud lifetime $\tau_{\rm life}$ as inputs.
 \item Lines 10-11: Define the local variables for node $n$. The set of children of $n$ is given by $\mathcal{C}_n$ and the set of parents is given by $\mathcal{P}_n$. As such, the numbers of children/parents at node $n$ are given by the sizes of the sets.
 \item Line 13: Calculate the number of MC outcomes at node $n$. This is equal to the number of child nodes (outgoing paths) if $n$ is a formation node ($\theta_{\rm child}(n) > \theta_{\rm par}(n)$), equal to the number of parent nodes (incoming paths) if $n$ is a destruction node ($\theta_{\rm child}(n) < \theta_{\rm par}(n)$), and equal to either quantity if $n$ is an intersection node ($\theta_{\rm child}(n) = \theta_{\rm par}(n)$). In general, it is therefore given by the maximum value of $\theta_{\rm child}$ and $\theta_{\rm par}$.
 \item Line 15: Calculate the number of MC outcomes that result in path termination at $n$. This is equal to zero if $n$ is a formation node or an intersection node, and equal to the reduction in the node number, $\theta_{\rm par}(n) - \theta_{\rm child}(n)$, if $n$ is a destruction node. In general, it is therefore given by the reduction in node number at any node, with a lower limit of zero.
 \item Lines 16-18: Use the random number $r_n$ for node $n$ to choose the first path taken at node $n$ by the first trajectory in the loop over $f \in \mathcal{F}$ to access $n$. The index of the outcome is $k$.
 \item Line 19: Cycle the path taken according to how many times node $n$ has already been accessed. For example, if node $n$ has $N_{\rm outcomes} = 3$ possible MC outcomes and has already been accessed $I_n=1$ time and taken the outcome $k = 2$, then the outcome is updated as $k = (2+1) \mod 3 = 0$.
 \item Line 20: Update the number of times that node $n$ has been accessed, for the next iteration.
 \item Lines 21-22: For a destruction node, the first $N_{\rm term}$ outcomes are designated as cloud destructions. The path/recursion is terminated and we return the cloud lifetime $\tau_{\rm life}$. For any other node type, $N_{\rm term} = 0$ and so this option is not accessed.
 \item Lines 23-25: If the path has not been terminated in the preceding if-clause, proceed to the $(k-N_{\rm term})$th child node of $n$ by continuing the recursion on this node. Iterate the cloud lifetime by the time interval $\Delta t$ and pass both arguments back to the start of the function.
\end{itemize}

\begin{figure}
 \label{Fig::schematic-MCpaths}
  \includegraphics[width=\linewidth]{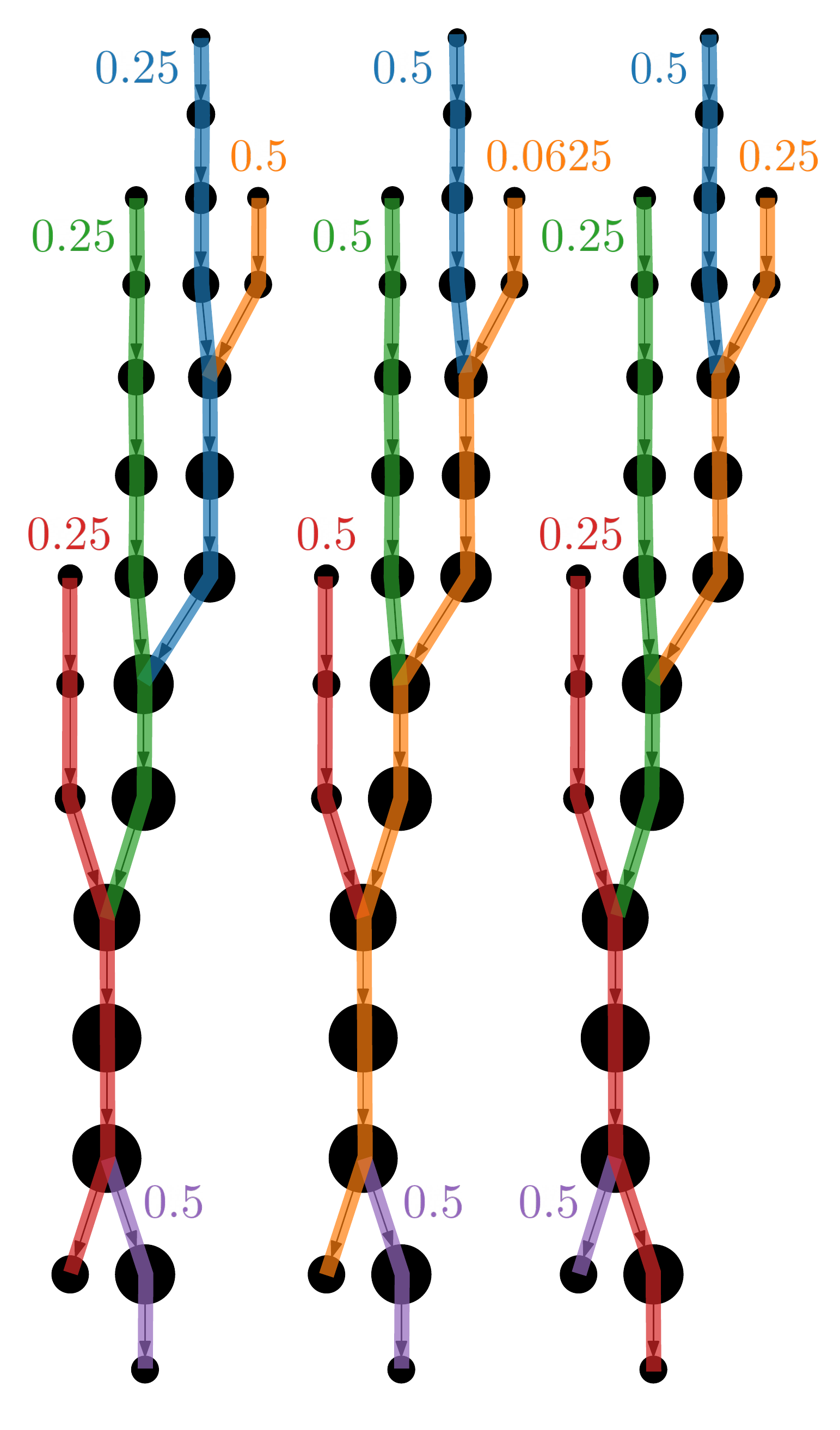}
  \caption{Schematic illustrating $3/16$ unique Monte Carlo realisations of trajectories through one connected component of the FLAT cloud history graph, obtained via the application of Algorithm~\ref{alg::cloud-lifetime}. The probability of obtaining each trajectory, relative to the case of a perfectly-straight path (no mergers or splits) is given by the number at each formation node. The illustration emphasises that a trajectory becomes exponentially less-likely as it passes through more mergers and splits. There are fewer Monte Carlo realisations containing such paths, although all of the Monte Carlo realisations (including the three depicted here) are equally-likely.}
\end{figure}

\begin{figure}
 \label{Fig::schematic-nodetypes}
  \includegraphics[width=\linewidth]{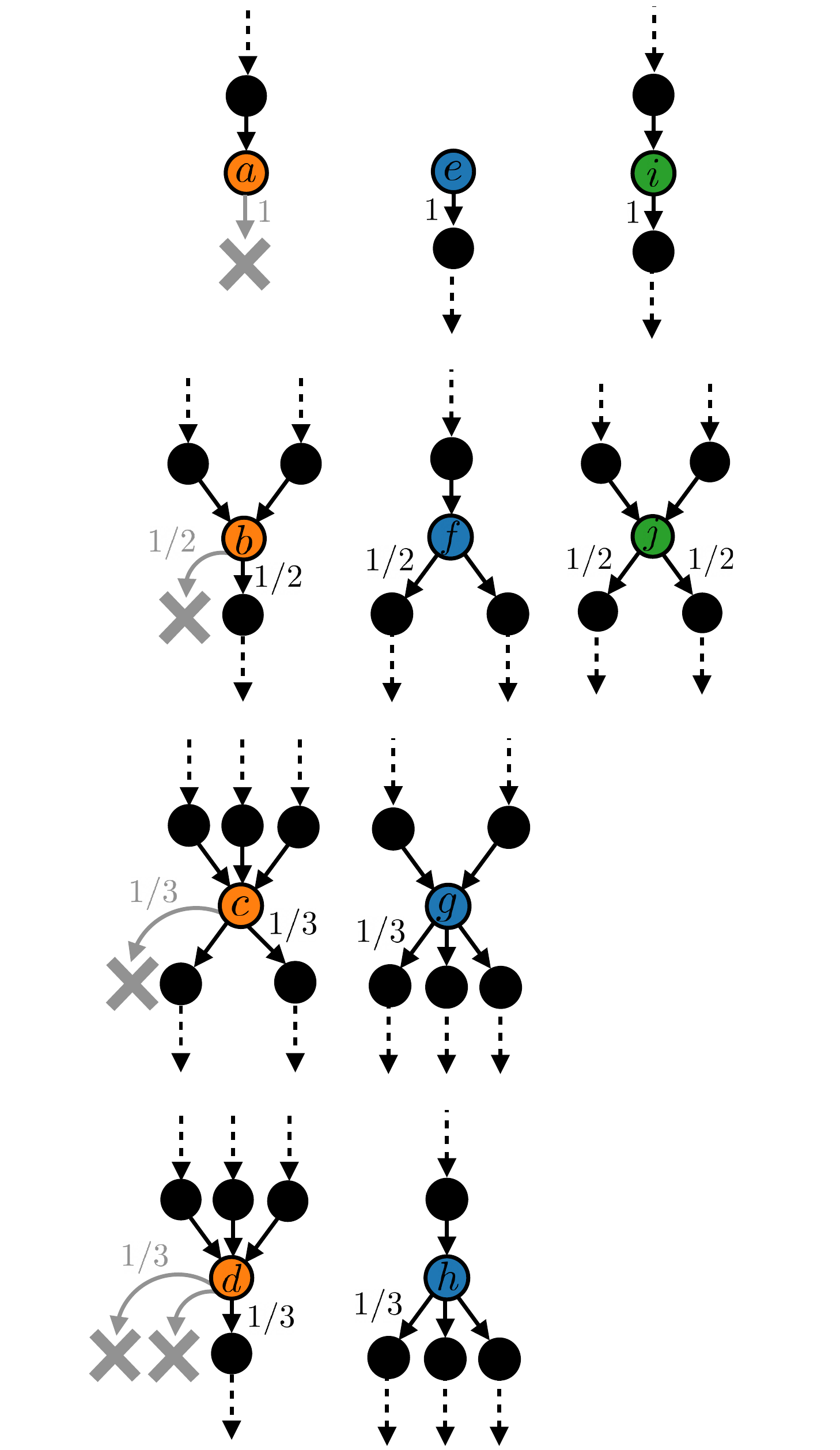}
  \caption{Schematic illustrating the MC outcomes at different types of node in the cloud evolution network. Destruction nodes are coloured orange (nodes $a$-$d$), formation nodes are coloured blue ($e$-$h$), and nodes that generate no net change in cloud number are coloured green (nodes $i$-$j$). The probabilities of the different MC outcomes in each case are given by the arrow labels. In the case of destruction nodes, a fraction $(\theta_{\rm par} - \theta_{\rm child})/\theta_{\rm par}$ of the total MC outcomes result in termination of the cloud evolutionary path, illustrated by grey crosses.}
\end{figure}

\label{lastpage}

\end{document}